\documentclass{aa}  

\usepackage{graphicx}
\usepackage{subfigure}
\usepackage{float}
\usepackage{amsmath}    
\usepackage{mathrsfs}    
\usepackage{bm}
\usepackage{amssymb}    
\usepackage{diagbox}    
\usepackage{enumerate}
\usepackage{float}
\usepackage{ulem}
\usepackage{url}
\usepackage{hyperref}
\usepackage{multirow}
\usepackage{multicol}
\usepackage{longtable}

\usepackage{txfonts}
\begin{document} 


  \title{A possibly solar metallicity atmosphere escaping from HAT-P-32b revealed by H$\alpha$ and He  absorption}  
  

   \author{Dongdong Yan\inst{1,}\inst{2,}\inst{3} 
                        \and
          Jianheng Guo \inst{1,}\inst{2,} \inst{3,} \inst{4}
           \and
          Kwang-il Seon \inst{5,}\inst{6}
          \and
          Manuel L\'{o}pez-Puertas \inst{7}
          \and
          Stefan Czesla \inst{8}
          \and
          Manuel Lamp\'{o}n \inst{7}
          }

   \institute{Yunnan Observatories, Chinese Academy of Sciences, P.O. Box 110, Kunming 650216, People's Republic of China \label{inst1}\\
              \email{yand@ynao.ac.cn}
         \and
          Key Laboratory for the Structure and Evolution of Celestial Objects, CAS, Kunming 650216, People's Republic of China \label{inst2}         
         \and
         International Centre of Supernovae, Yunnan Key Laboratory, Kunming 650216, China \label{inst3} 
         \and
         School of Astronomy and Space Science, University of Chinese Academy of Sciences, Beijing, People's Republic of China\label{inst4} 
        \and             
         Korea Astronomy \& Space Science Institute, 776 Daedeokdae-ro,
Yuseong-gu, Daejeon 34055, Republic of Korea \label{inst5} 
         \and
         Astronomy and Space Science Major, University of Science and
Technology, 217, Gajeong-ro, Yuseong-gu, Daejeon 34113, Republic of Korea \label{inst6} 
         \and
         Instituto de Astrofísica de Andaluc\'{i}a (IAA-CSIC), Glorieta de la Astronom\'{i}a s/n, 18008 Granda, Spain\label{inst7}
         \and
         Thüringer Landessternwarte Tautenburg, Sternwarte 5, D-07778 Tautenburg, Germany \label{inst8}
     }

   \date{Received October 10, 2023; accepted March 25, 2024}

 
  \abstract
   {This paper presents a hydrodynamic simulation that couples detailed non-local thermodynamic equilibrium (NLTE) calculations of the hydrogen and helium level populations to model the H$\alpha$ and He 10830 transmission spectra of the hot Jupiter HAT-P-32b. A Monte Carlo simulation is applied to calculate the number of Ly$\alpha$ resonance scatterings, which is the main process for populating H(2). 
In the examined parameter space, only the models with H/He $\geq$ 99.5/0.5, $(0.5 \sim 3.0)$ times the fiducial value of $F_{\rm XUV}$, $\beta_m = 0.16\sim 0.3$, can explain the H$\alpha$ and He 10830 lines simultaneously. We find a mass-loss rate of $\sim (1.0\sim 3.1) \times 10^{13}$ g s$^{-1}$, consistent with previous studies. Moreover, we find that the stellar Ly$\alpha$ flux should be as high as $4 \times 10^{5}$ erg cm$^{-2}$ s$^{-1}$, indicating high stellar activity during the observation epoch of the two absorption lines.
 Despite the fact that the metallicity in the lower atmosphere of HAT-P-32b may be super-solar, our simulations tentatively suggest it is close to solar in the upper atmosphere. The difference in metallicity between the lower and upper atmospheres is essential for future atmospheric characterisations.}
   
   {}
   {
}
   {
   }
   %
   {
   }


   \maketitle
%

\section{Introduction}
\label{sec:1}
Hydrogen and helium are the two most abundant elements in the atmospheres of giant planets both in solar and extrasolar systems. During the last two decades, escaping atmospheres of hydrogen and helium have been detected by the excess absorption lines through transmission spectroscopy. To detect hydrogen during transit, the Ly$\alpha$ line in the ultraviolet has been initially employed because of its high absorption depth caused by the predominant presence of hydrogen in the ground state H(1s) \citep{2003Natur.422..143V,2010A&A...514A..72L,2012A&A...543L...4L,2014ApJ...786..132K,2015Natur.522..459E,2018A&A...620A.147B,2022NatAs...6..141B,2022AJ....163...68Z}. As a complementary information, the H$\alpha$ line in the optical is able to probe planetary hydrogen atoms in the excited state H(2) \citep{2012ApJ...751...86J,2018AJ....156..154J,2018NatAs...2..714Y,2021A&A...645A..22Y,2018A&A...616A.151C,2019AJ....157...69C,2021AJ....161..152C,2020MNRAS.494..363C,2020A&A...635A.171C,2021A&A...645A..24B,2022AA...657A...6C}. 
To detect helium, its infrared triplet (hereafter He 10830), which arises from the transition between 2$^3$S and 2$^3$P state, is utilized. Excess absorption of He 10830 by the planet's atmosphere during transit is a signature of the existence of metastable helium H(2$^3$S) \citep{2000ApJ...537..916S,2018ApJ...855L..11O,2018Natur.557...68S,2018A&A...620A..97S,2018Sci...362.1388N,2018Sci...362.1384A,2019A&A...629A.110A,2022AJ....164...24K,2022AJ....163...67Z,2023AJ....165...62Z,2023SciA....9F8736Z}. Because both the H$\alpha$ and He 10830 lines are not contaminated by interstellar absorption and can be observed by ground-based telescopes with large apertures and high spectral resolutions, it is common to use them to probe hydrogen and helium in planetary atmospheres.

The excess absorptions of both the H$\alpha$ and He 10830 lines have been detected only in four systems, i.e., HD 189733b \citep{2012ApJ...751...86J,2016MNRAS.462.1012B,2017AJ....153..217C,2018A&A...620A..97S,2020A&A...639A..49G,2022AJ....164..237Z,2023A&A...677A.164A}, WASP-52b \citep{2020A&A...635A.171C,2020AJ....159..278V,2022AJ....164...24K,2022AJ....164..234V,2023A&A...677A.164A}, HAT-P-32b \citep{2022AA...657A...6C,2023SciA....9F8736Z}, and HAT-P-67b \citep{2023AJ....166...69B,2023arXiv230708959G}. Simultaneously modelling the transmission spectra of H$\alpha$ and He 10830 lines can help to constrain the physical parameters of the planetary atmosphere. In particular, the mass-loss rate of exoplanets with an expanding atmosphere has been found to be as high as 10$^{9}$-10$^{13}$ g s$^{-1}$ \citep{2011ApJ...733...98G, 2013ApJ...766..102G, 2016A&A...586A..75S,2021MNRAS.507.3626K,2022ApJ...929...52K}; such a high mass-loss rate could play a crucial role in planetary compositions and dynamics, and especially in the evolutions and architectures (e.g., occurrence) of small sized exoplanets \citep{2018AJ....156..264F,2018MNRAS.479.5012O,2020ApJ...888L...5Y,2022AJ....164..234V, 2023A&A...673A.140L}. The energy-limited approach allows to estimate the planetary mass-loss rate based on the absorbed energy \citep{2003ApJ...598L.121L,2007A&A...472..329E}. However, for planets with very high gravitational potentials or exposed to intense X-ray and extreme ultraviolet (XUV) flux, this approach can yield a mass-loss rate significantly different from that obtained using more complex, self-consistent models (\citep{2016A&A...586A..75S,2019ApJ...880...90Y,2021A&A...648L...7L, 2021A&A...655A..30C}. Moreover, the energy-limited approach cannot provide detailed atmospheric structures that are necessary to interpret the transmission signals. Therefore, a self-consistent hydrodynamic calculation is essential for analysing the observations and gaining information about the escaping atmosphere.

In atmospheric modelling, some physical parameters or quantities are challenging to measure or estimate. For example, the XUV ($1-912 \rm\AA$) radiation from host stars, which plays an essential role in the heating and photochemistry of the planetary atmosphere, is difficult to measure because the extreme ultraviolet (EUV) radiation is readily absorbed by the interstellar medium. There are some works to reconstruct the XUV spectra. For instance, ``MUSCLES Treasury Survey'' is dedicated to reconstruct the spectral energy distributions (SEDs) of M and K type stars in a range of $5 \rm\AA$ to 5.5 micron, including the XUV band \citep{2016ApJ...820...89F}. The X-ray spectra of these stars can be detected using the Chandra X-ray Observatory and XMM-Newton instruments or simulated using the APEC models \citep{2001ApJ...556L..91S}. The EUV spectra are obtained  by using either the empirical scaling relation based on Ly$\alpha$ flux \citep{2014ApJ...780...61L}, with the Ly$\alpha$ spectra reconstructed from model fits that take the stellar flux and interstellar medium into account \citep{2016ApJ...824..101Y}, or through the use of the differential emission measure models \citep{2021ApJ...913...40D}.
For the late-type stars, \cite{2011A&A...532A...6S} derived a relation between the X-ray and EUV flux, and both of these can be estimated for a given stellar age. A few works obtain the SEDs by using the XSPEC software \citep{1996ASPC..101...17A}. However, despite employing these various methods, obtaining
detailed XUV SEDs remains challenging. All the reconstructed XUV spectra of extrasolar systems are rather uncertain, both in flux and in spectral shape. While the He 10830 absorption is sensitive to the XUV flux \citep{2022ApJ...936..177Y}, these order-of-magnitude uncertainties on the XUV spectra do not necessarily translate to order-of-magnitude uncertainties on the absorption signatures, as suggested by \cite{2022A&A...667A..54L}.
Modelling the observed transmission spectral lines can help constrain the XUV radiation \citep{2021ApJ...907L..47Y,2022ApJ...936..177Y}.  
In addition, the hydrogen-to-helium abundance ratio, H/He, may be significantly different from that of the Sun.
By modelling the He 10830 transmission spectra of some exoplanets, it was found that a  much higher H/He, or equivalently a much lower helium abundance, is estimated in the planetary upper atmosphere \citep{2020A&A...636A..13L,2021A&A...647A.129L,2023A&A...673A.140L,2021MNRAS.500.1404S,2022ApJ...936..177Y,2022ApJ...927..238R, 2022A&A...658A.136F}. Why does this happen? Is helium hard to escape compared to hydrogen, or, is the origin of helium different on this planet from that on the Sun? More detailed studies of the escaping helium atmosphere are required to obtain clues on such questions.

The first work that could explain both the H$\alpha$ and He 10830 transmission spectra in an exoplanet system was done by \cite{2022AA...657A...6C}, who used two independent models to interpret the H$\alpha$ and He 10830 signals of HAT-P-32b. 
In the work of \cite{2022ApJ...936..177Y} (hereafter Paper I), we simultaneously modelled the H$\alpha$ and He 10830 absorption of WASP-52b and fitted the observation quite well by using a hydrodynamic model coupled with a non-local thermodynamic model. To calculate the H(2) population, we performed a Monte Carlo simulation and calculated the Ly$\alpha$ intensity inside the planetary atmosphere. 
The model that reproduced both lines could constrain the XUV flux ($F_{\rm XUV}$) and SEDs, as well as the hydrogen-to-helium abundance ratio H/He, and finally help estimate the mass-loss rate of the planet. In that work, the stellar Ly$\alpha$ photons were the main source to populate H(2), while the Ly$\alpha$ photons produced in the planetary atmosphere were negligible. Is this result universal? To answer this question, it is worthwhile to analyse more systems.

Using the CARMENES spectrograph, \cite{2022AA...657A...6C} detected pronounced, time-dependent absorption in the H$\alpha$ and He 10830 triplet lines with maximum depths of about 3.3 \% and 5.3 \%, respectively, and they attributed these absorptions to the planetary atmosphere. In addition, an early ingress of redshifted absorption was observed in both lines.
\cite{2022AA...657A...6C} performed pioneering work on modelling the absorption spectra of H$\alpha$ and He 10830 of HAT-P-32b. To explain the H$\alpha$ transmission spectrum, they used a 1D hydrodynamic model including a non-local thermodynamic equilibrium (NLTE) treatment of hydrogen level populations, using the model of \cite{2019ApJ...884L..43G}. Their model only considers the species of hydrogen and electrons. To estimate the XUV heating, they assumed that the XUV flux received at the sub-stellar point represents the flux illuminated across the whole planetary surface.
The mass-loss rate is estimated by $\pi\rho u r^2$ to account for the average effect of stellar heating, where $\rho$ is the density of the atmosphere, $u$ is the velocity, and $r$ is the distance from the planetary center. In their work, the H(2) population that causes the H$\alpha$ absorption, is formed in a narrow layer around $1.8 R_P$. 
To model the He 10830 transmission spectrum, \cite{2022AA...657A...6C} used a variation of the 1D isothermal Parker wind model (see \cite{2020A&A...636A..13L}), which assumes a constant speed of sound. Different from that of \cite{2019ApJ...884L..43G}, this model takes into account both hydrogen and helium. In the model, the temperature, hydrogen-to-helium abundance ratio H/He, and mass-loss rate are free parameters. By comparing the model transmission spectrum of He 10830 with the observation, they found that H/He can be either 90/10 or 99/1. A mass-loss rate of about $10^{13}$ g s$^{-1}$ was obtained by modelling the H$\alpha$ and He 10830 lines in \cite{2022AA...657A...6C}.
Recently, \cite{2023A&A...673A.140L} reanalysed the upper atmosphere of HAT-P-32b, and showed that this planet undergoes photoevaporation at a mass-loss rate of about (1.30$\pm$0.7)$\times 10^{13}$g s$^{-1}$, and the atmosphere temperature is in the range of 12,400$\pm$2900 K. They also constrained the H/He ratio in the planetary upper atmosphere to be about 99/1. 

Recently, using high-resolution spectroscopy of He 10830 obtained from the Hobby-Eberly Telescope, \cite{2023SciA....9F8736Z} detected the escaping helium associated with giant tidal tails of HAT-P-32b. The He 10830 absorption depth at mid-transit was $\sim$ 8.2\%, about 1.5 times higher than that observed by \cite{2022AA...657A...6C}. The variability of He 10830 absorption in HAT-P-32b may imply a variation of stellar activity of HAT-P-32.
To explain the asymmetric signals, \cite{2023SciA....9F8736Z} used a 3D hydrodynamic simulation and predicted the Roche lobe overflow with extended tails. They estimated a mass-loss rate of about 1.07$\times 10^{12}\rm g \ s^{-1}$, which is about an order of magnitude lower than the results of \cite{2022AA...657A...6C} and \cite{2023A&A...673A.140L} (see Sec. 4.3).

We note that \cite{2022AA...657A...6C} analysed the H$\alpha$ and He 10830 transmission spectra separately with two independent models. In contrast, in this paper, we intend to simultaneously model the H$\alpha$ and He 10830 transmission spectra of HAT-P-32b using a self-consistent model for characterizing its upper atmosphere. 
Since both the H$\alpha$ and He 10830 absorptions were detected at the same time in \cite{2022AA...657A...6C}, we mainly compare our model with their observation. In addition, we also discuss the He 10830 observation by \cite{2023SciA....9F8736Z}.  
Unlike the models used by \cite{2022AA...657A...6C}, we simulate Ly$\alpha$ resonance scattering by assuming that both the stellar and planetary atmospheres are spherical. In addition, to calculate the population of He(2$^3$S), we use the atmospheric structure obtained from a non-isothermal hydrodynamic model. 
Hence, this work presents an important modelling effort to understand the
atmospheric outflow of HAT-P-32b and provides additional constrains on its upper atmosphere.

The paper is organized as follows.
In Section \ref{sec:Methods}, we describe the method. Section \ref{sec:Result-part1} compares the results with observations and other modelling works. In section \ref{sec: discuss}, we discuss some relevant subjects to the present work, including the escape of helium. Finally, Section \ref{sec: consclusion} summarizes our work and presents our conclusions.

\section{Method}\label{sec:Methods}

\begin{figure}
	\begin{minipage}[t]{0.5\textwidth}
		\centering
		\includegraphics[width=\textwidth]{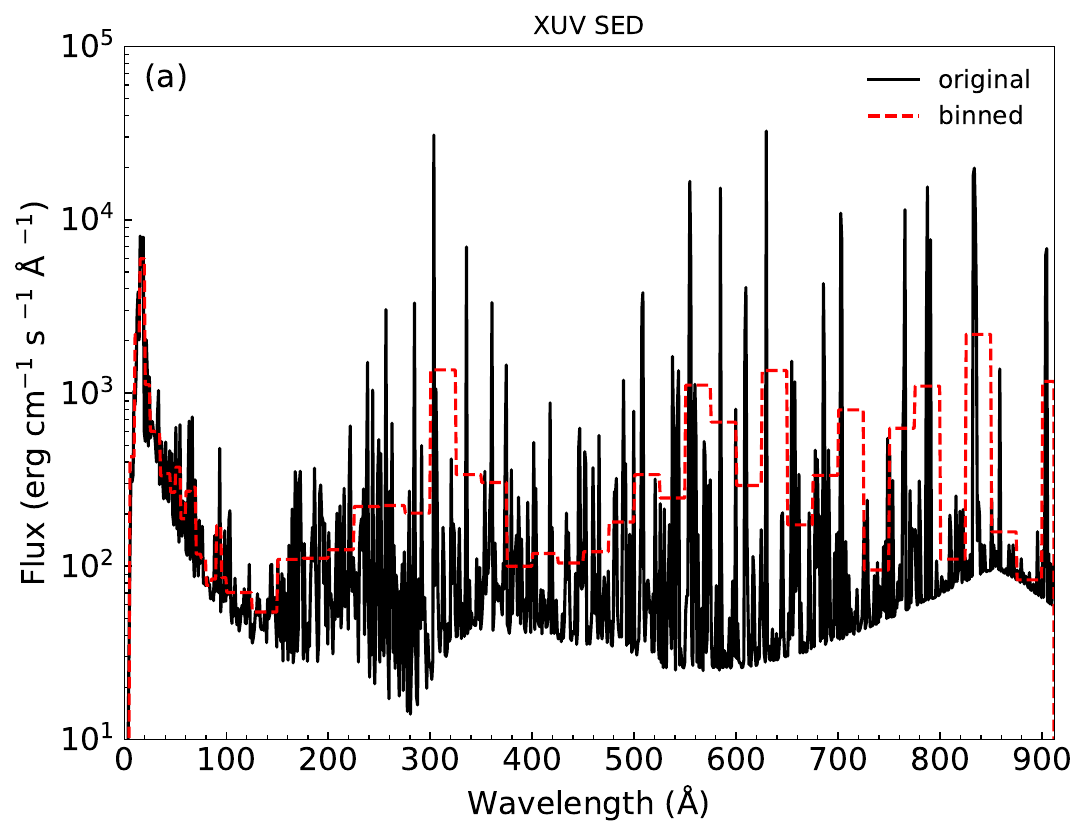}
	\end{minipage}
	\begin{minipage}[t]{0.5\textwidth}
		\centering
		\includegraphics[width=\textwidth]{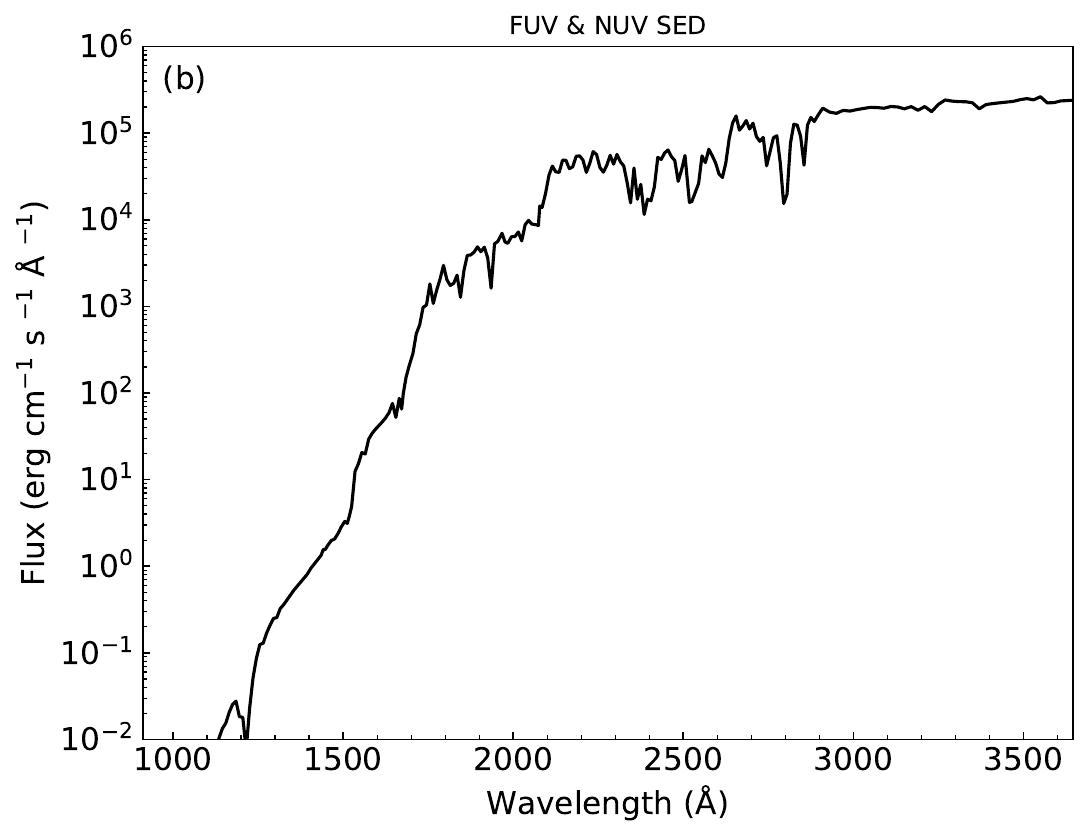}
	\end{minipage}
    \caption{The stellar XUV, FUV, and NUV SEDs. Panel (a) shows the original spectrum taken from Sanz-Forcada (in prep), which is the same as that used in \cite{2022AA...657A...6C} and \cite{2023A&A...673A.140L}. The red dashed line represents the binned spectrum according to our model settings. Panel (b) shows the  SED in FUV and NUV wavelengths, which was reconstructed based on the stellar atmosphere model of \cite{2003IAUS..210P.A20C}. }
    \label{fig:xuv_nuv}
\end{figure}

We used the 1D hydrodynamic model \citep{2016ApJ...818..107G,2018ChA&A..42...81Y,2019ApJ...880...90Y} to simulate the atmospheric structure of HAT-P-32b and to obtain the atmospheric temperature, velocity, and particle number densities. The planetary and stellar parameters are based on reported observations \citep{2011ApJ...742...59H,2022AA...657A...6C}. HAT-P-32 is an F-type star with $M_\star = 1.160 M_\sun$ and $R_\star = 1.219 R_\sun$. HAT-P-32b is a hot Jupiter with $M_P = 0.585 M_J$ and $R_P = 1.789 R_J$.
The equilibrium temperature is 1805 K, which is adopted as the temperature at the lower boundary in our model. 
The integrated flux in the XUV band is a crucial parameter in the simulations. 
We adopted here the spectral energy distribution (SED) up to 920 Å and the luminosity in the ranges of 100-504 $\rm\AA$ and 100-920 $\rm\AA$ of \cite{2022AA...657A...6C}, which are based on XMM-Newton X-ray observations. The fluxes in each band along with other model parameters, are presented in Table \ref{tab:paramaters}.
Using the SED, the total XUV flux is estimated to be about $ F_{\rm XUV} = 4.2\times 10^{5} \rm erg$ cm$^{-2}$ s$^{-1}$ (hereafter $F_0$, the fiducial XUV flux) at the planetary orbit (0.0343 AU). Note that this value is at the sub-stellar point and there is no flux on the nightside for the 1D model. Therefore,
in our calculation this value is divided by a factor of 4, which accounts for the uniform redistribution of the stellar radiation energy around the planet.
The SED index, $\beta_m = F(1-100\rm\AA)$/$F(1-912\rm\AA$) as defined in \cite{2021ApJ...907L..47Y}, is about 0.16. We reconstructed the SEDs in the FUV and NUV (912- 3646 $\rm\AA$) wavelengths based on the stellar atmosphere model of \cite{2003IAUS..210P.A20C}. The stellar XUV, FUV, and NUV SEDs are shown in Figure \ref{fig:xuv_nuv}.

The chemical composition of HAT-P-32b is initially assumed to be the same as that of HAT-P-32, which is identical to the solar abundances except for [Fe/H] = -0.04. This can lead to a slight change in the atmospheric scale height compared to using solar metallicity. In this study, we have not included the cooling effects caused by metals (C, N, O, Si, and their respective ions).
Cooling by metals like Fe II or Mg II can be significant under certain circumstances \citep{2022AJ....163...67Z, 2023ApJ...951..123H}. For example, \cite{2022AJ....163...67Z} found that Fe II compromises 30\% of the cooling from (1.6-2.0) $R_P$ and (2.9-3.4) $R_P$ in the atmosphere with a solar metallicity of TOI 560.01, but cooling by hydrogen is still dominant; for a higher metallicity, the cooling of metals becomes more important. \cite{2023ApJ...951..123H} found that cooling by Mg II in the atmosphere of WASP-121b can be dominant at altitude lower than $\sim 1.4 \ R_P$.
However, these metals are not included in our model, thus their cooling effects are not considered. We defer such studies to a future work. Initially, we assume a number ratio of hydrogen to helium to be the same as the solar value (H/He = 92/8, \citealt{2009ARA&A..47..481A}). We then explore the effects of varying the H/He ratio while maintaining constant metallicity.
We note that while the stellar metallicity is almost solar, the planetary atmosphere may have a different metallicity. Transmission spectroscopy of the lower atmosphere conducted with Hubble Space Telescope and Spitzer Infrared Array Camera photometry, suggested that the atmosphere of HAT-P-32b is metal-rich, with a  metallicity of probably exceeding 100 (or 200) times the solar value, i.e., $log(Z/Z_{\sun}) = 2.41 ^{+0.06}_{-0.07}$ \citep{2020AJ....160...51A}. However, the upper atmosphere remains unclear whether such a high metallicity could appear, as heavy species may not escape as easily as the light ones. Multi-fluid hydrodynamic simulations have shown that the decoupling of heavy and light species can occur when collisions between them become less frequent. In such cases, the heavy species would tend to remain in the lower atmosphere, while light ones rise to the upper atmosphere and then eventually escape \citep{2019ApJ...872...99G, 2023ApJ...953..166X}. In this work, to investigate the influence of metallicity, we examine the cases of Z = 1, 10, 30, 50, 100, 200, and 300  times solar metallicity, following the convention of \cite{2022AJ....163...67Z}. Note that for the case of the solar metallicity (Z =1 ), the planetary atmosphere consists of 73.80\% hydrogen, 24.85\% helium, and 1.34\% metals by mass \citep{2009ARA&A..47..481A}. 
We assume an iron abundance of [Fe/H] = -0.04 in the Z = 1 case. The mass fractions of H, He, and metals are 31.72\%, 10.68\%, and 57.6\% for Z = 100, respectively, and 14.74\%, 4.96\%, and 80.30\% for Z = 300. In our fiducial models (see Table \ref{tab:paramaters} for details) with Z > 1, the hydrogen-to-helium abundance ratio remains 92/8. For 10 < Z < 50, only the fiducial models are considered. We calculate only a few models by varying other parameters ($ F_{\rm XUV}$ and $\beta_m$) for each metallicity when Z > 100 because of computational expense.

In the simulations, the pressure at the lower boundary ($1 \ R_P$) is 1 $\mu$bar. The upper boundary is 10 $R_P$, which is larger than the radius of the host star (about 6.7 $R_P$). 
The Ly$\alpha$ cooling and stellar tidal force are also considered in the simulations.

Since the hydrodynamic model cannot calculate the level populations of H(2) and H(2$^3$S), we solve the independent equations of non-local thermodynamic rate equilibrium of H(2) and H(2$^3$S) to obtain the populations (see Paper I for details). 
The number density of the H($2p$) state is primarily determined by the number of Ly$\alpha$ pumping events and thus by the Ly$\alpha$ radiation within the atmosphere \citep{2013ApJ...772..144C, 2017ApJ...851..150H, 2021ApJ...907L..47Y,2022ApJ...936..177Y,2023ApJ...951..123H}. Thus, detailed simulations of Ly$\alpha$ radiative transfer are done by using LaRT \citep{2020ApJS..250....9S, 2022ApJS..259....3S, 2022ApJ...936..177Y}.
The incident stellar Ly$\alpha$ photons and those generated within the planetary atmosphere are the two sources of Ly$\alpha$ in this model. 
The planetary Ly$\alpha$ source is due to the collisional excitation and recombination, as shown in Equation (8) of Paper I \citep{2017ApJ...851..150H, 2022ApJ...936..177Y}. 
For the incident stellar Ly$\alpha$, we adopt a similar line profile as in Paper I due to the lack of Ly$\alpha$ observations for this system. In other words, we assume a double Gaussian line profile of Ly$\alpha$ at the outer boundary of the planetary atmosphere, with a width of 49 kms$^{-1}$ centered at $\pm$74 kms$^{-1}$.
\cite{2013ApJ...766...69L} presented Ly$\alpha$ fluxes at a distance of
1 AU (in erg cm$^{-2}$ s$^{-1}$) for four F-type stars: Procyon with Ly$\alpha$ flux of 77.1, HR 4657 of 27.8 , $\zeta$ Dor of 46.5 and $\xi$ Her of 22.0. Here, we choose a moderate value of 46.5 ($\zeta$ Dor) for HAT-P-32b and adjust it for its orbital distance. The resulting Ly$\alpha$ flux is 39,524 erg cm$^{-2}$ s$^{-1}$ (hereafter referred to as $F_{\rm Ly\alpha_0}$, representing the fiducial value of Ly$\alpha$ flux $F_{\rm Ly\alpha}$). 
To find the best fit to the observations, we also explore a wide parameter space of $F_{\rm XUV}$, $\beta_m$, H/He, and $F_{\rm Ly\alpha}$. All the parameters and models in this work are listed in Table \ref{tab:paramaters}.

\section{Results}\label{sec:Result-part1}

In this section, we present the modelled He 10830 and H$\alpha$ transmission spectra and compare them with the observations of \cite{2022AA...657A...6C}. The wavelengths of the He 10830 triplet are 10829.09, 10830.25, and 10830.33 $\rm\AA$. In our model, the two latter lines are considered as a single merged line centered at 10830.29$\rm\AA$ because they are practically unresolved \citep{1971PhRvA...3..908D}. We use the wavelengths in air and the wavelength range of 10828.2$\rm\AA$-10831.8$\rm\AA$ is considered. The line center of H$\alpha$ is at 6562.8 $\rm\AA$ in air, and the simulations and analyses are performed in the wavelength range of [6561.0, 6564.6 $\rm\AA$].
In this work, we shift the observed He 10830 and H$\alpha$ data in vacuum to that in air for model comparison.

After simulating the He 10830 and H$\alpha$ transmission spectra, we calculate the $\chi^2$. 
The data contain structures, such as the spikes in the H$\alpha$ line, that cannot be reproduced by our model and are likely attributable to systematic noise. In the calculation of our $\chi^2$ value,  we, therefore, applied the procedure suggested by \cite{2023A&A...677A.164A}, which renormalizes $\chi^2$ by dividing by the lowest obtained $\chi^2$ value. This is tantamount to enlarging the error bars, thereby accounting for systematic noise contributions.
After binning the data, we have a number of degree of freedom of 52, which leads to a standard deviation of the renormalized reduced $\chi^2$ distribution of $\sigma = \sqrt{(2 \times 52)}/52 \sim 0.2$. 
Then we define models with $\chi^2 < 1.6$ (within 3$\sigma$) as good-fit models of the observation.

\subsection{Modelling He 10830 and H$\alpha$ transmission spectra with solar metallicity (Z = 1)}\label{sec:Result-part1-5}

\subsubsection{H/He = 92/8}\label{sec:Result-part1-5}
\begin{figure}
	\begin{minipage}[t]{0.5\textwidth}
		\centering
		\includegraphics[width=\textwidth]{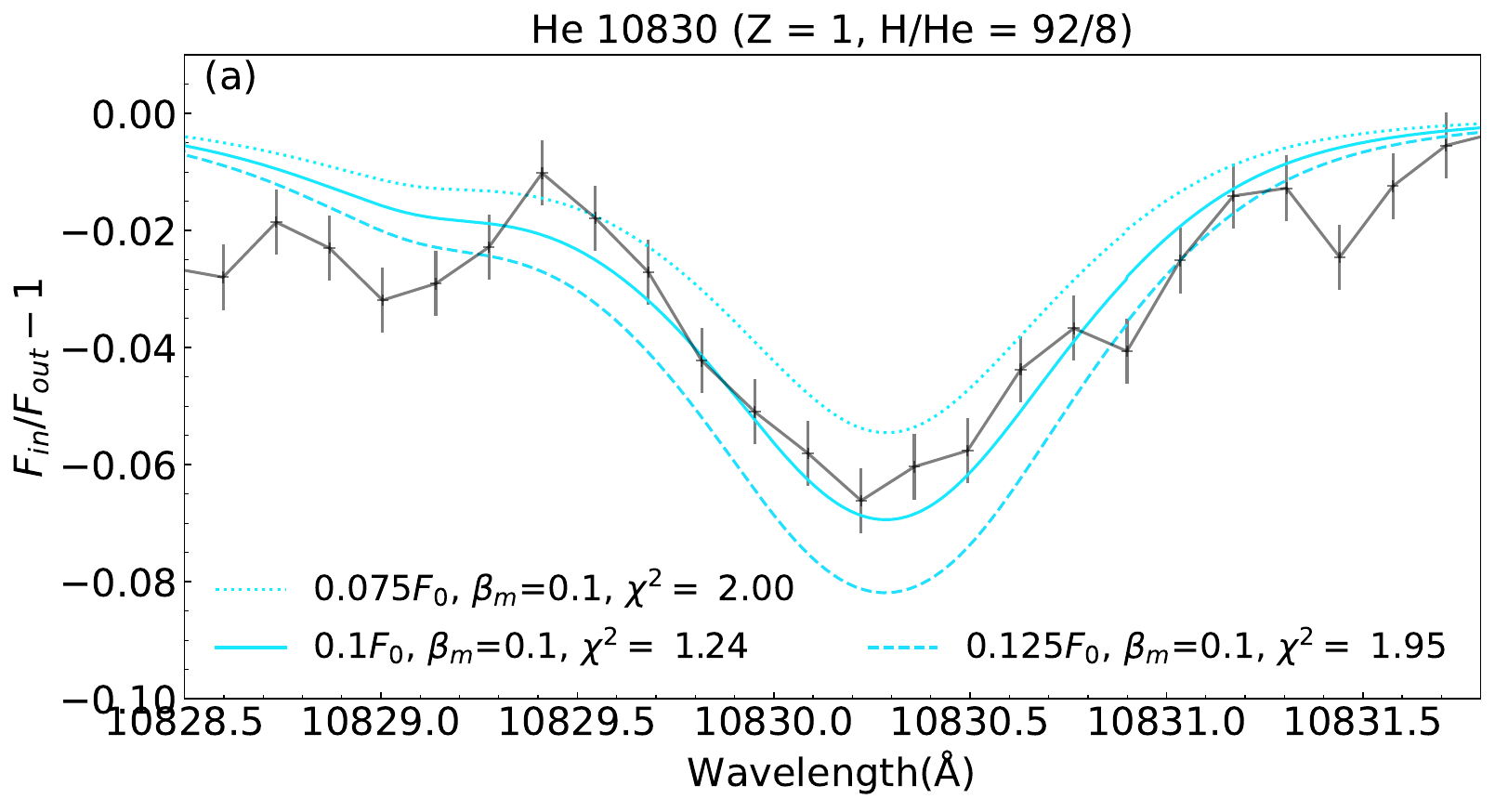}
	\end{minipage}
	\begin{minipage}[t]{0.5\textwidth}
		\centering
		\includegraphics[width=\textwidth]{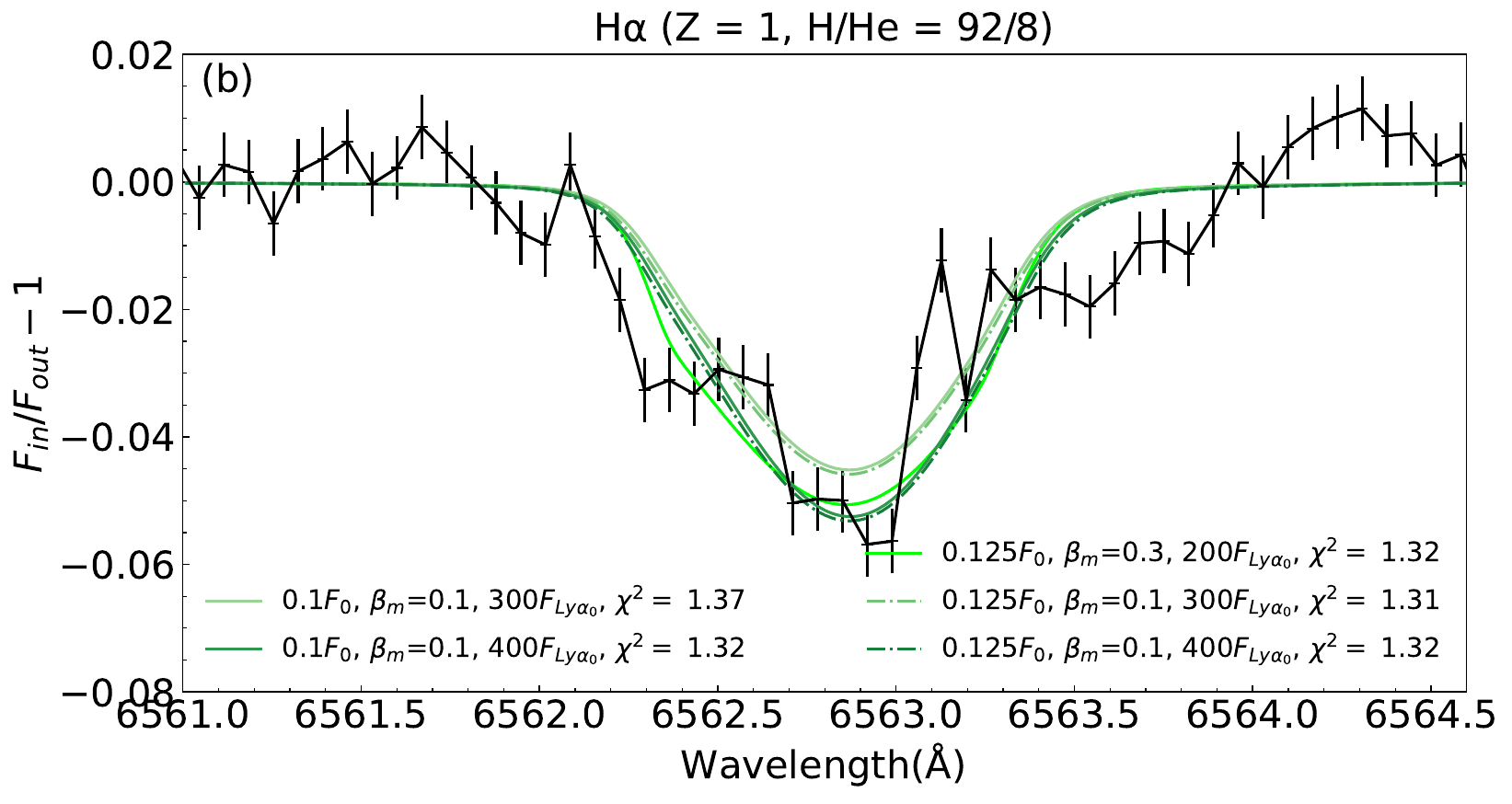}
	\end{minipage}
    \caption{Comparison of He 10830 and H$\alpha$ transmission spectra with the models that reproduce both the lines well with solar metallicity (Z = 1) and H/He ratio (92/8).
    In each panel, the label in the y-axis $F_{in}/F_{out} -1 $ is the in-transit flux over out-of transit flux minus 1, which represents the relative absorption depth. The legends show the model parameters and the values of $\chi^2$. The lines with error bars are the observation data from \cite{2022AA...657A...6C}.} 
    \label{model_best_fit_Z1_H92}
\end{figure}

Figure \ref{model_best_fit_Z1_H92} (a) shows the the transmission spectra of He 10830 for models with solar metallicity (Z = 1) and the fiducial hydrogen-to-helium abundance ratio (H/He = 92/8). We find that the He 10830 absorption is very sensitive to the change of $F_{\rm XUV}$. A high  $F_{\rm XUV}$ tends to lead to a  high absorption. In the explored parameter space of $F_{\rm XUV} \geq  0.125 F_0$ and $\beta_m$, the model for $F_{\rm XUV}= 0.125 F_0$ and $\beta_m = 0.1$ gives $\chi^2 = 1.95$, but still larger than 1.6 (the 3$\sigma$ limit). One can see that the modelled absorption depth is obviously higher than the observation in a wide wavelength range.  We then slightly decrease $F_{\rm XUV}$ to 0.1$F_0$ and 0.075 $F_0$. As $F_{\rm XUV}$ decreases to 0.1$F_0$, $\chi^2$ decreases, then increases as it further decreases to 0.075 $F_0$. Therefore, the model of $F_{\rm XUV}= 0.1 F_0$ and $\beta_m = 0.1$ gives the lowest $\chi^2$ value in the parameter space and can explain the He 10830 observation well.

In Figure \ref{model_best_fit_Z1_H92} (b) we show models that can explain the H$\alpha$ line with $F_{\rm XUV}= 0.1 \ (0.125) F_0$ and $\beta_m = 0.1$. We find that, unlike He 10830, changing the XUV flux from 0.1$F_0$ to 0.125$F_0$ does not result in a significant change of the H$\alpha$ absorption depth, as can be seen in the dash-dotted lines in Figure \ref{model_best_fit_Z1_H92} (b). 
However, in this case, a stellar Ly$\alpha$ flux as high as 300-400 times the fiducial value is required ($F_{\rm Ly\alpha}/F_{\rm Ly\alpha_0} \approx 300-400$). This corresponds to (1.2-1.6)$\times 10^7$ erg cm$^{-2}$ s$^{-1}$, which is about 0.5\% of the bolometric flux received from the star ($\sim 2.39 \times 10^9$ erg cm$^{-2}$ s$^{-1}$). \cite{2013ApJ...766...69L} investigated the ratio of Ly$\alpha$ flux to X-ray flux for stars with different stellar types, establishing an upper limit below 100 (approximately 60-80) and a lower limit of around 0.6 for F5V-G9V stars. 
In our work, the fiducial X-ray luminosity is about 2.3$\times 10 ^{29}$ erg s$^{-1}$, so the fiducial X-ray flux at the planetary orbit $F_{\rm X_0}$ is about 6.95 $\times 10^4$ erg cm$^{-2}$ s$^{-1}$. For the model of $F_{\rm XUV} = 0.1 F_0$, the X-ray flux $F_{\rm X} = 0.1 F_{\rm X_0} $ = 6.95 $\times 10^3$ erg cm$^{-2}$ s$^{-1}$. Therefore, $F_{\rm Ly\alpha}/F_{\rm X} \sim (1700-2300)$, which is one order of magnitude higher than the upper limit in \cite{2013ApJ...766...69L}. Increasing the XUV flux $F_{\rm XUV}$ and spectral index $\beta_m$ can mitigate the requirement of such a high stellar Ly$\alpha$ flux. For instance, the lime solid line in Figure \ref{model_best_fit_Z1_H92} (b) shows that for the model with $F_{\rm XUV} = 0.125 F_0$, when $\beta_m$ increases to 0.3, we obtained a reduction of $F_{\rm Ly\alpha}/F_{\rm Ly\alpha_0} \approx 200$ to fit the absorption line.
 However, low  $F_{\rm XUV}$ and $\beta_m$ are preferred for explaining the He 10830 observation.

\subsubsection{H/He = 99/1, 99.5/0.5, and 99.9/0.1}\label{sec:Result-part1-5}

The reason very low $F_{\rm XUV}$ and $\beta_m$ are needed for models with H/He = 92/8 to explain the He 10830 observation is that a much lower population of metastable helium is necessary in the upper atmosphere. In this section, we attempt to explain the observations by reducing the helium abundance, i.e., increasing H/He to 99/1, 99.5/0.5, or 99.9/0.1.

\begin{figure*}
    \centering
    \subfigure{\includegraphics[width=0.49\textwidth]{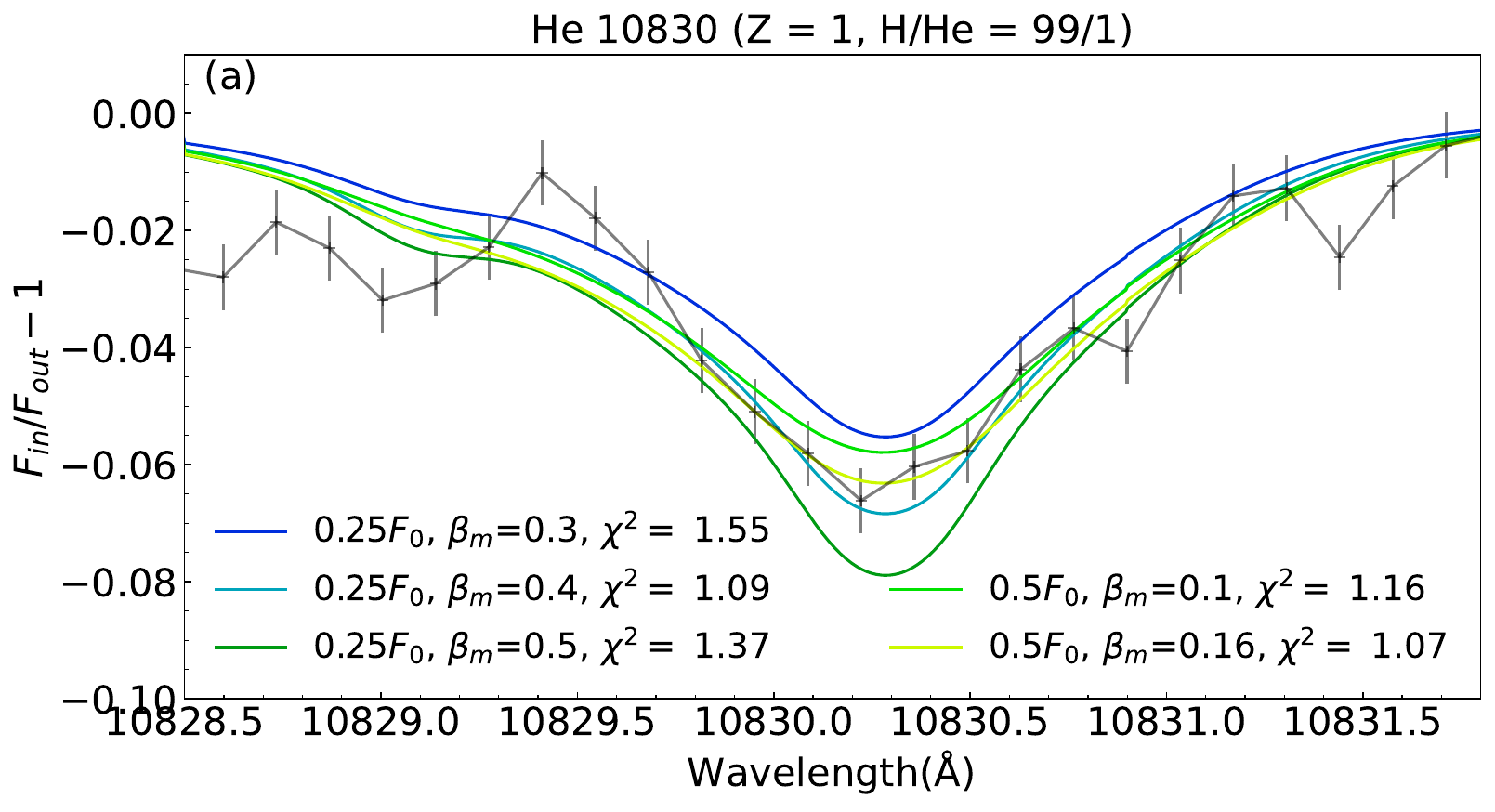}} 
    \subfigure{\includegraphics[width=0.49\textwidth]{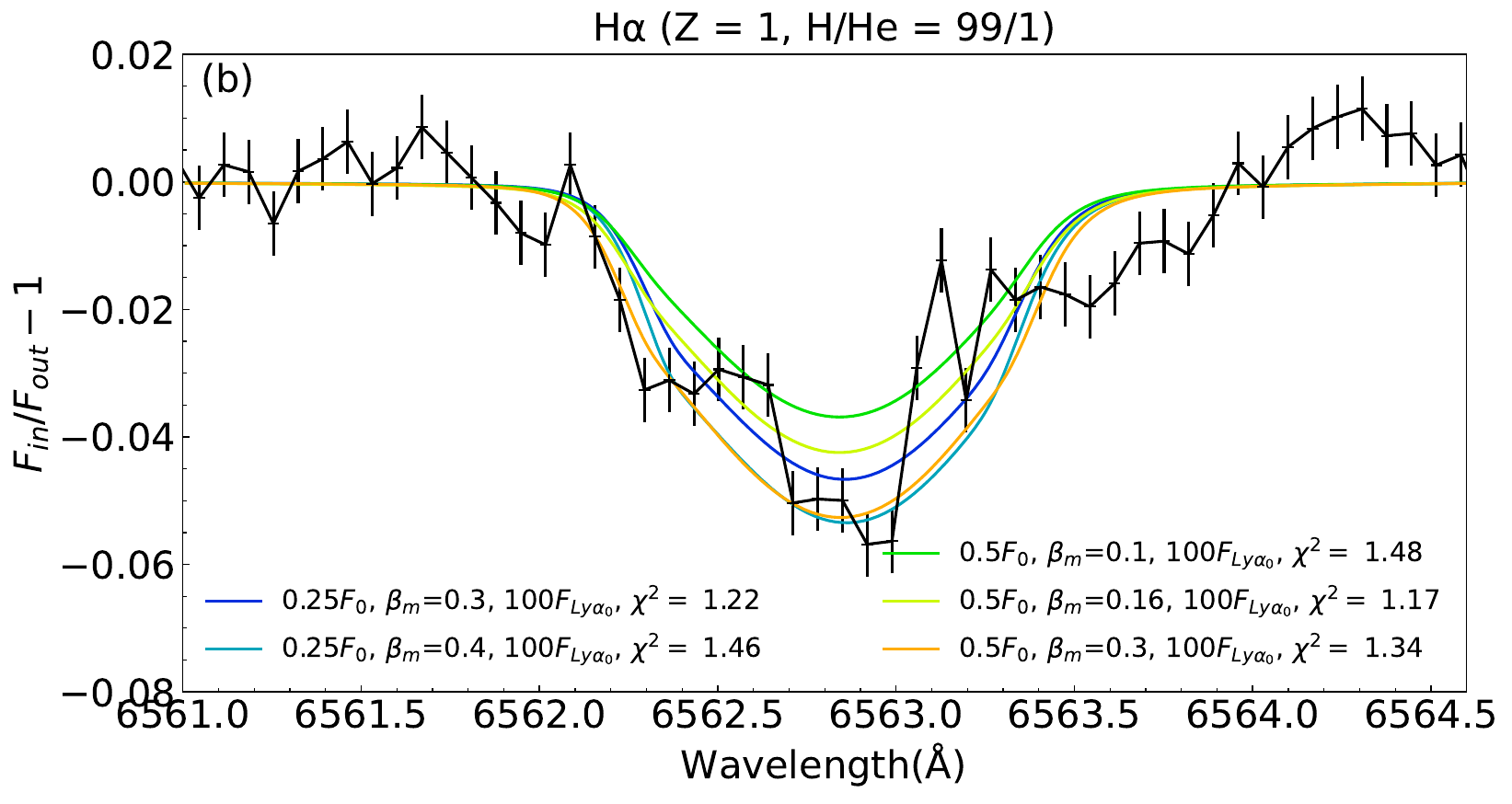}}
    \\ 
    \centering
    \subfigure{\includegraphics[width=0.49\textwidth]{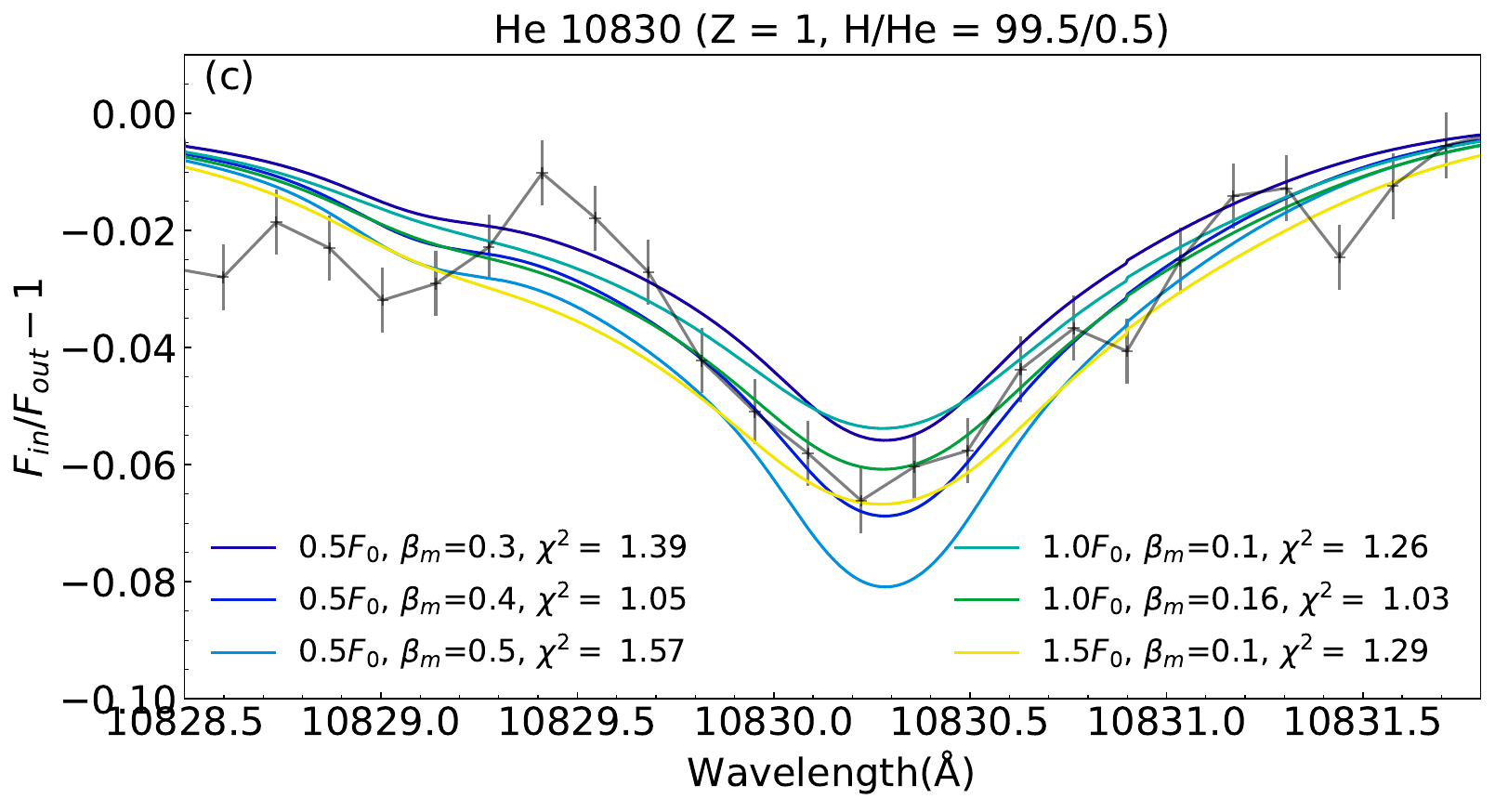}}
    \subfigure{\includegraphics[width=0.49\textwidth]{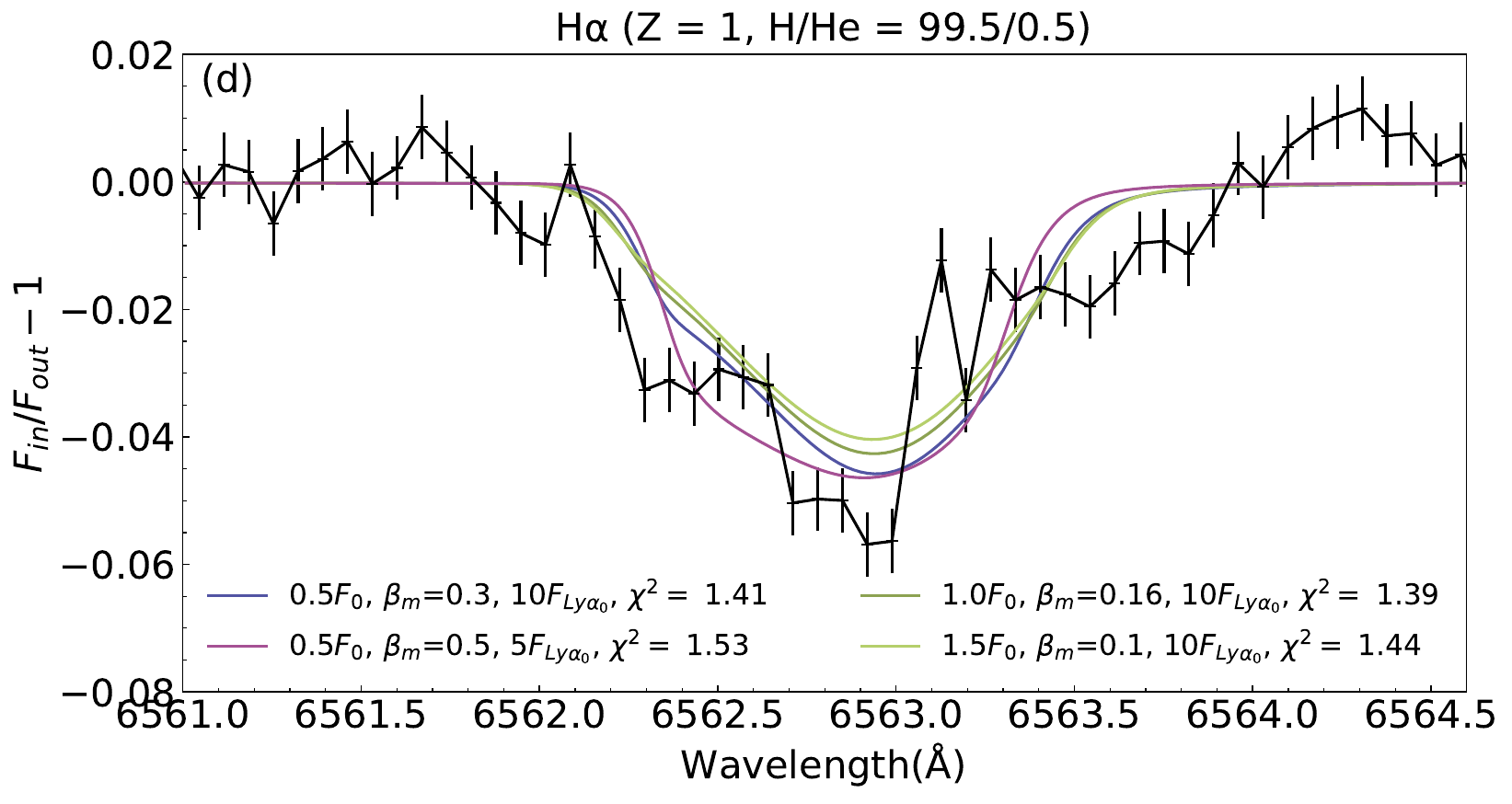}}
        \\ 
    \centering
    \subfigure{\includegraphics[width=0.49\textwidth]{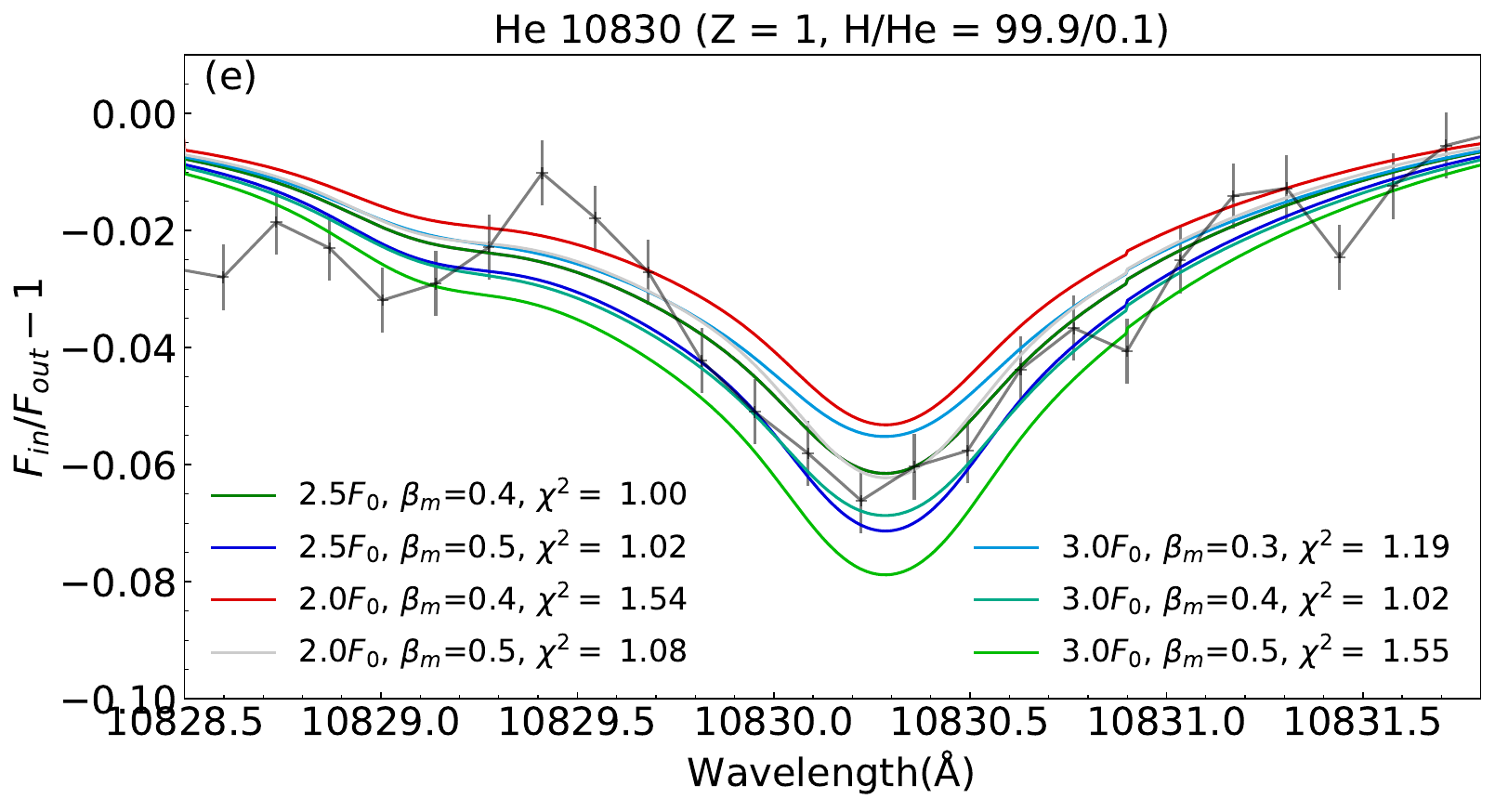}}
    \subfigure{\includegraphics[width=0.49\textwidth]{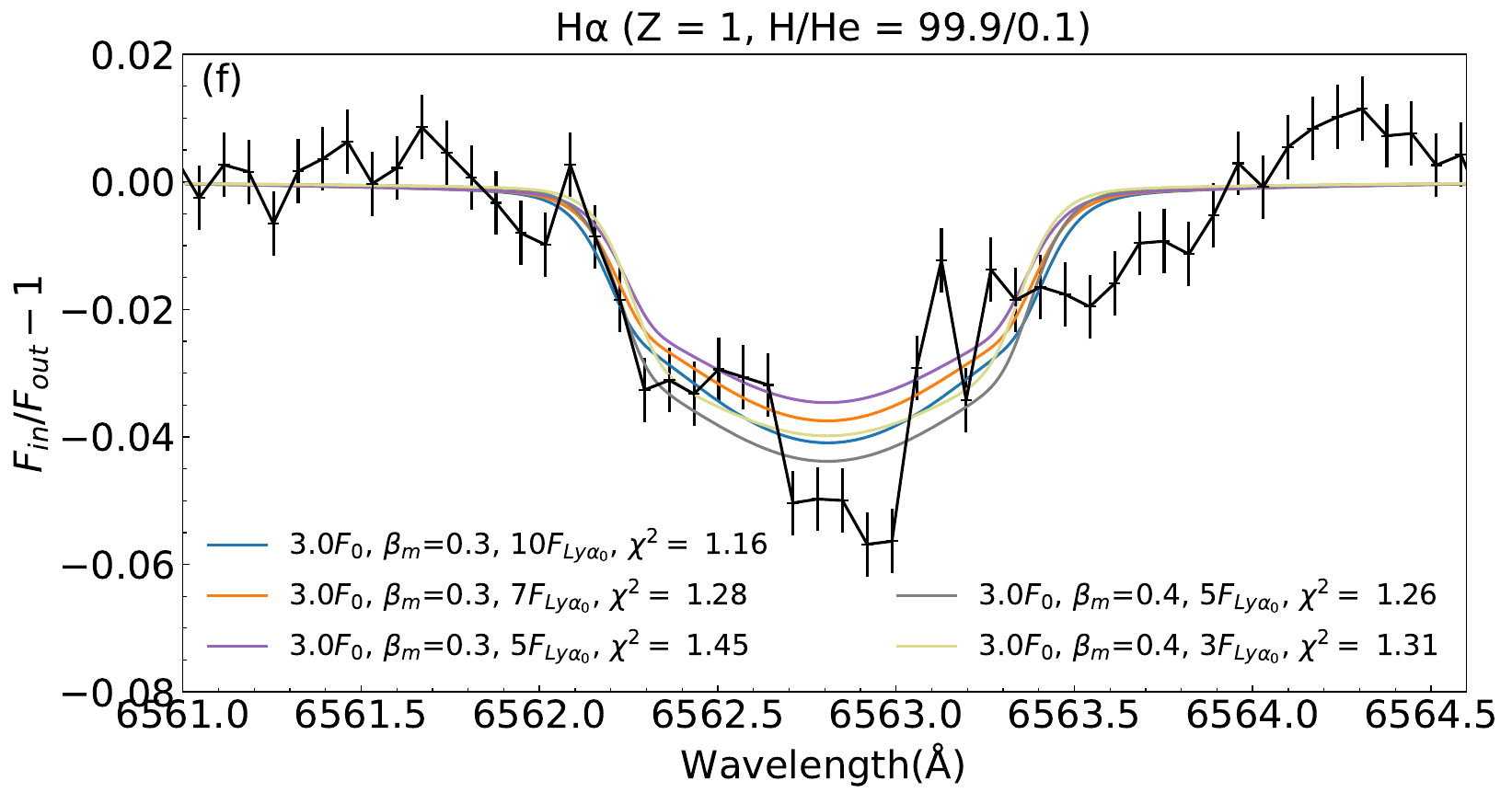}}
    \caption{The He 10830 and H$\alpha$ transmission spectra of the models with H/He = 99/1, 99.5/0.5, and 99.9/0.1 that fit the observation. The lines with error bars are the observation data from \cite{2022AA...657A...6C}.} 
    \label{fig:He10830-models-all}
\end{figure*}

Figure \ref{fig:He10830-models-all} shows the models for various H/He ratios that can fit both the He 10830 and H$\alpha$ observations. Here, for every H/He case, we initially identify models that can fit the He 10830 observation, and subsequently select from these models those can fit the H$\alpha$ observation. Note that there could be additional models that match the H$\alpha$ observation, but they are not included in the plot because they do not align with the He 10830 observation. Although we present the results with $\beta_m \geq 0.4 $, it is noteworthy that this value of $\beta_m$ may be unrealistically high, as such a spectral index $\beta_m \geq 0.4 $ is not commonly seen for F-type stars \citep{2011A&A...532A...6S}. Therefore, we aim to identify and define models with $\beta_m < 0.4 $ as good fits for both lines.

We found that the models with high $F_{\rm XUV}$ and $\beta_m$ tend to lead to high absorption levels. The hydrogen-to-helium abundance ratio also affects the He 10830 absorption significantly; increasing the H/He ratio, an increase of $F_{\rm XUV}$ and $\beta_m$ are necessary for explaining the observations. 
It can be seen that when H/He = 99/1, only the models with 
$ 0.25 F_0 \leq F_{\rm XUV} \leq 0.5 F_0$ can match the He 10830 observation. 
To explain the H$\alpha$ absorption, a stellar Ly$\alpha$ flux needs to be as high as 100 times the fiducial value. In this case, $F_{\rm Ly\alpha}/F_{\rm X} \sim (115-230)$, still a few times higher than the upper limit in \cite{2013ApJ...766...69L}.
For H/He = 99.5/0.5, models with $ 0.5 F_0 \leq F_{\rm XUV} \leq 1.5 F_0$ can match the He 10830 observation, and a stellar Ly$\alpha$ flux about 10 times the fiducial value
would be enough for fitting the H$\alpha$ observation. Therefore, $F_{\rm Ly\alpha}/F_{\rm X}$ decreases to about 4-12, which is well within the range of the values given in \cite{2013ApJ...766...69L}. $F_{\rm Ly\alpha} = 10 F_{\rm Ly\alpha_0}$ seems very high ($4 \times 10^{5}$ erg cm$^{-2}$ s$^{-1}$) , but it is still within a reasonable range.
$F_{\rm Ly\alpha_0}$ is a moderate value for the F-type stars in \cite{2013ApJ...766...69L}. 
During the X-ray observations of HAT-P-32b by Sanz-Forcada
et al. (in prep) and \cite{2022AA...657A...6C}, a flare was detected,
emphasizing the high activity of the star. Such high activity
implies that the stellar Ly$\alpha$ flux would also be high.
For example, WASP-121 was also an active star according to \cite{2023ApJ...951..123H}  and its $F_{\rm Ly\alpha}$ was inferred to be about $1 \times 10^{5}$ erg cm$^{-2}$ s$^{-1}$ at the planetary orbit. Therefore, the high $F_{\rm Ly\alpha}$ is justified for HAT-P-32.

However, for models with H/He = 99.9/0.1, $F_{\rm XUV} \geq 2.0 F_0$ and $\beta_m \geq 0.3 $ are needed. In particular, $\beta_m \geq 0.4 $ is required in most cases; $\beta_m = 0.3 $ only appears for $F_{\rm XUV} = 3.0 F_0$. In the models with $F_{\rm XUV} = 3.0 F_0$ and $\beta_m = 0.3 $, $F_{\rm Ly\alpha} = 5 F_{\rm Ly\alpha_0}$, which is almost equal to the X-ray flux, would be sufficient to fit the H$\alpha$ line absorption profile within 3$\sigma$; $F_{\rm Ly\alpha} = 10 F_{\rm Ly\alpha_0}$ can give a lower $\chi^2$, and thus a better fit.
Such high $F_{\rm XUV}$ and $\beta_m$ values could be possible if the host star HAT-P-32 had a high activity when the absorptions of H$\alpha$ and He 10830 were observed. Therefore, from the above analyses, it is likely that the H/He ratio is higher than 99/1 in the upper atmosphere of this planet.

\begin{figure*}
    \centering
    \subfigure{\includegraphics[width=0.5\textwidth]{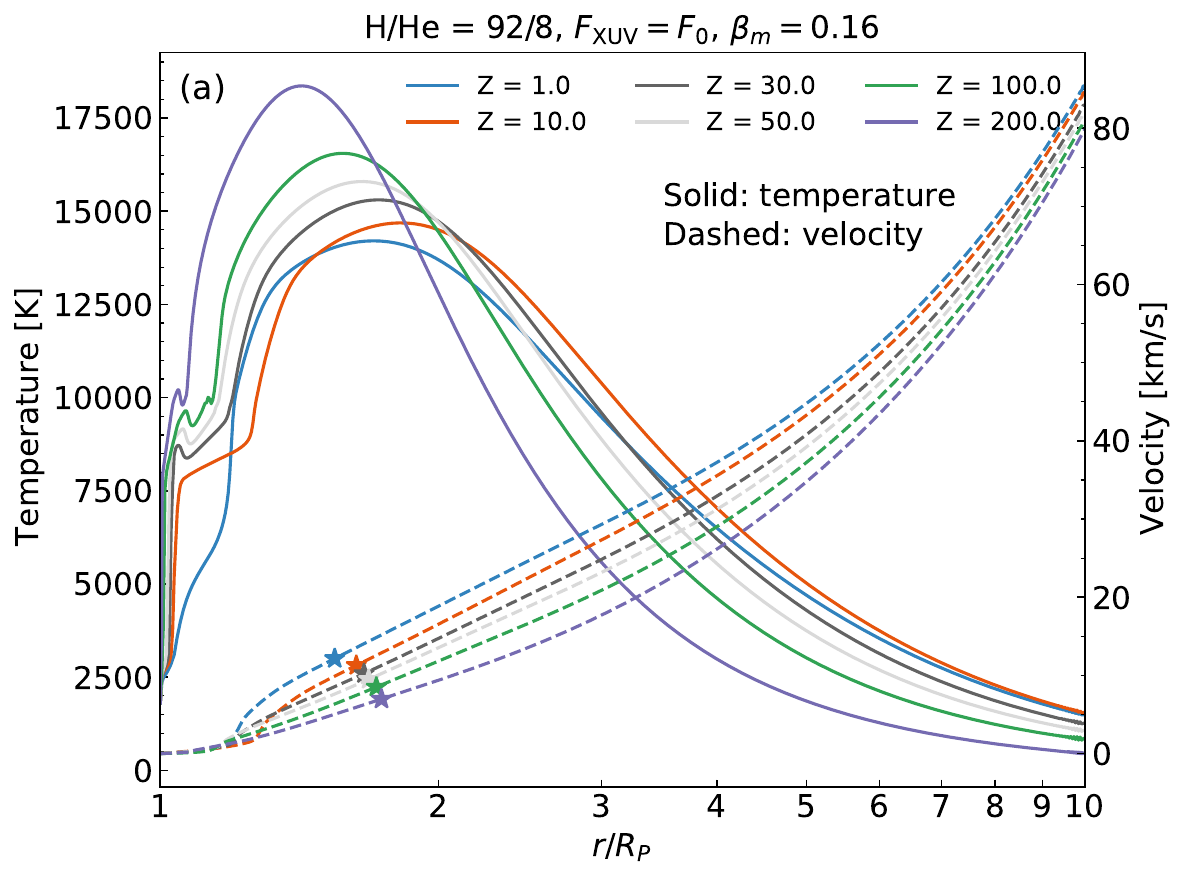}}
    \subfigure{\includegraphics[width=0.49\textwidth]{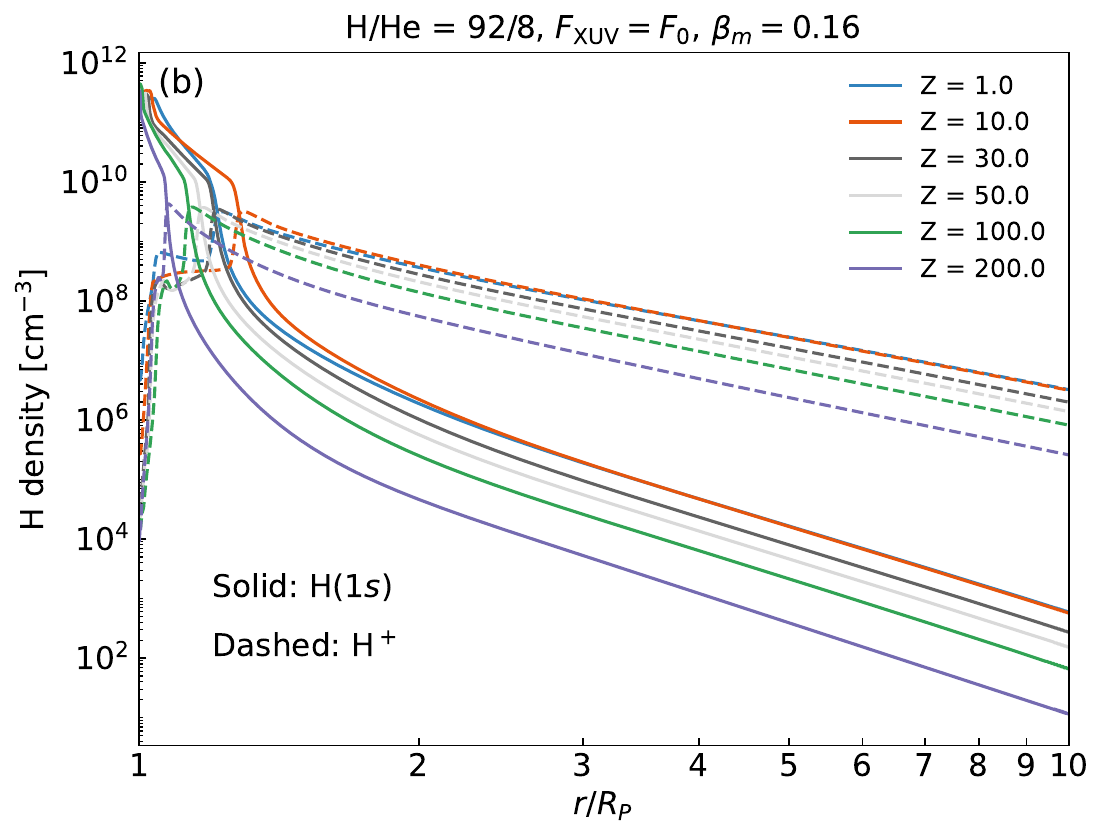}}
    \\ 
    \centering
    \subfigure{\includegraphics[width=0.49\textwidth]{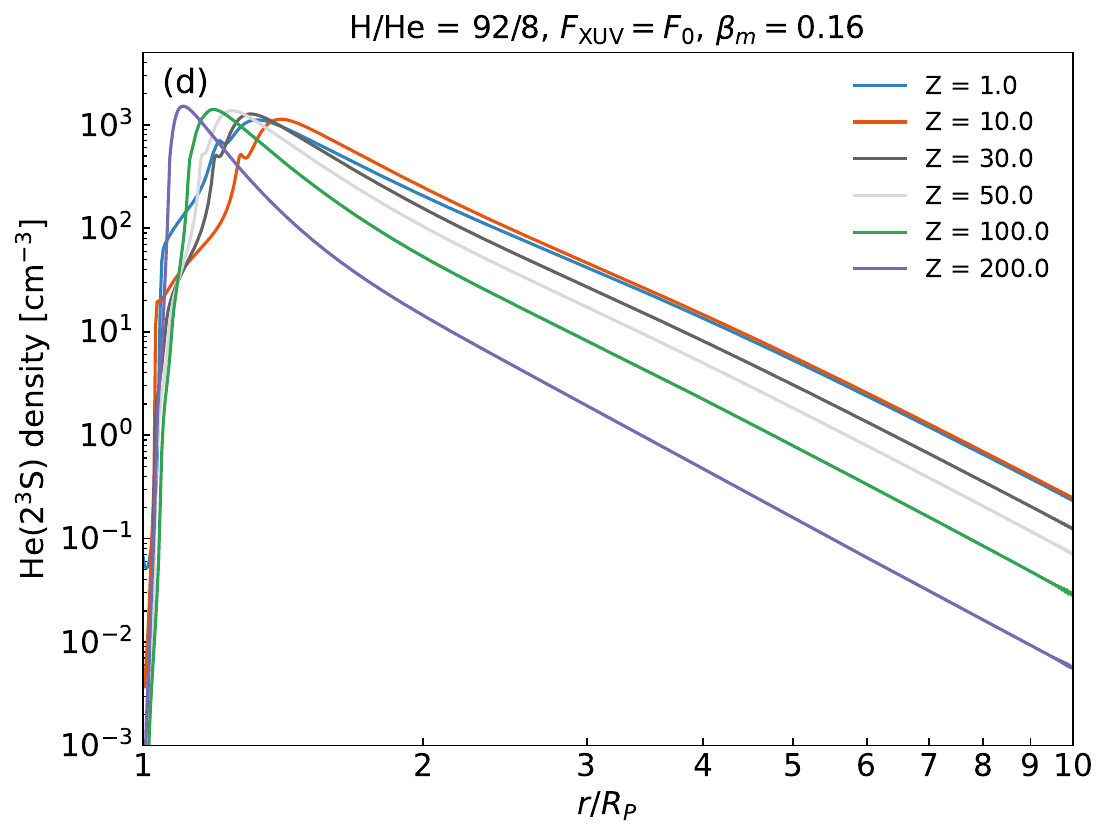}}
    \subfigure{\includegraphics[width=0.5\textwidth]{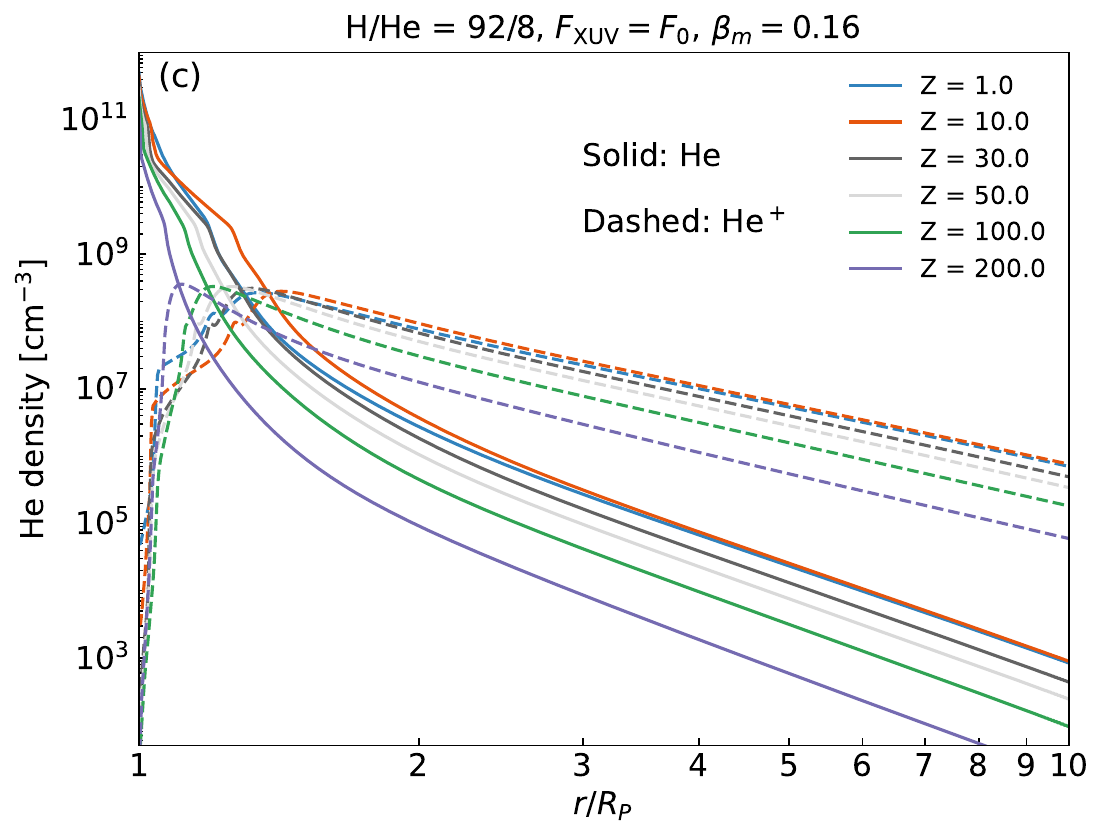}}
    \caption{The atmospheric structure of HAT-P-32b for models of H/He = 92/8, $F_{\rm XUV} = F_0$, and $\beta_m = 0.16$, but with different metallicities range from Z = 1 to Z = 200. (a) Temperature (denoted in solid lines) and velocity (dashed lines). The asterisks on the velocity lines represent the sonic points. (b) Number densities of hydrogen atoms H(1s) and ions H$^+$. (c) Number densities of helium atoms He and ions He$^+$. (d) Number densities of helium atoms in the metastable state, He(2$^3$S).} 
    \label{fig:atm_cases_z}
\end{figure*}

\begin{figure}
	\begin{minipage}[t]{0.5\textwidth}
		\centering
		\includegraphics[width=\textwidth]{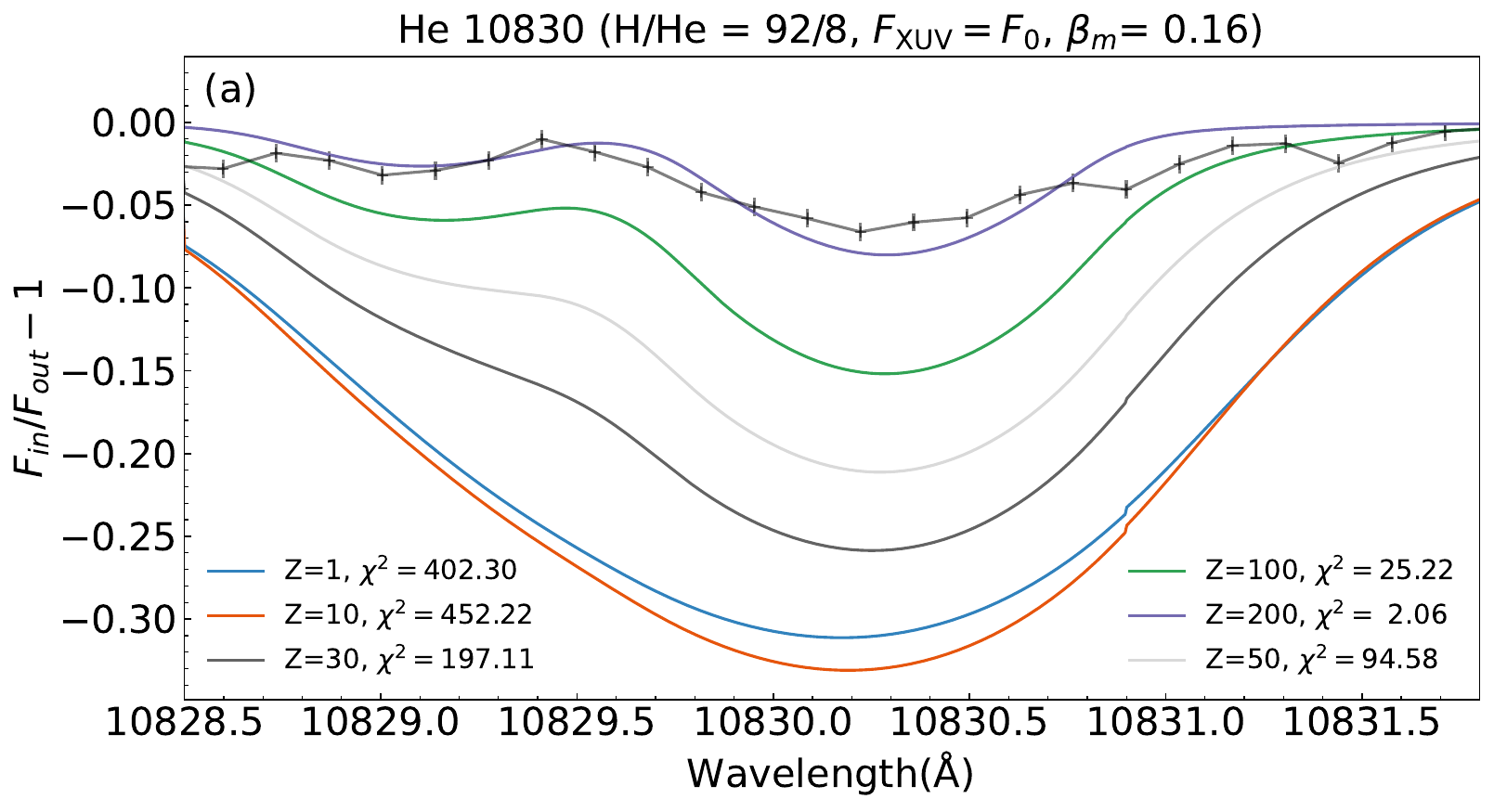}
	\end{minipage}
	\begin{minipage}[t]{0.5\textwidth}
		\centering
		\includegraphics[width=\textwidth]{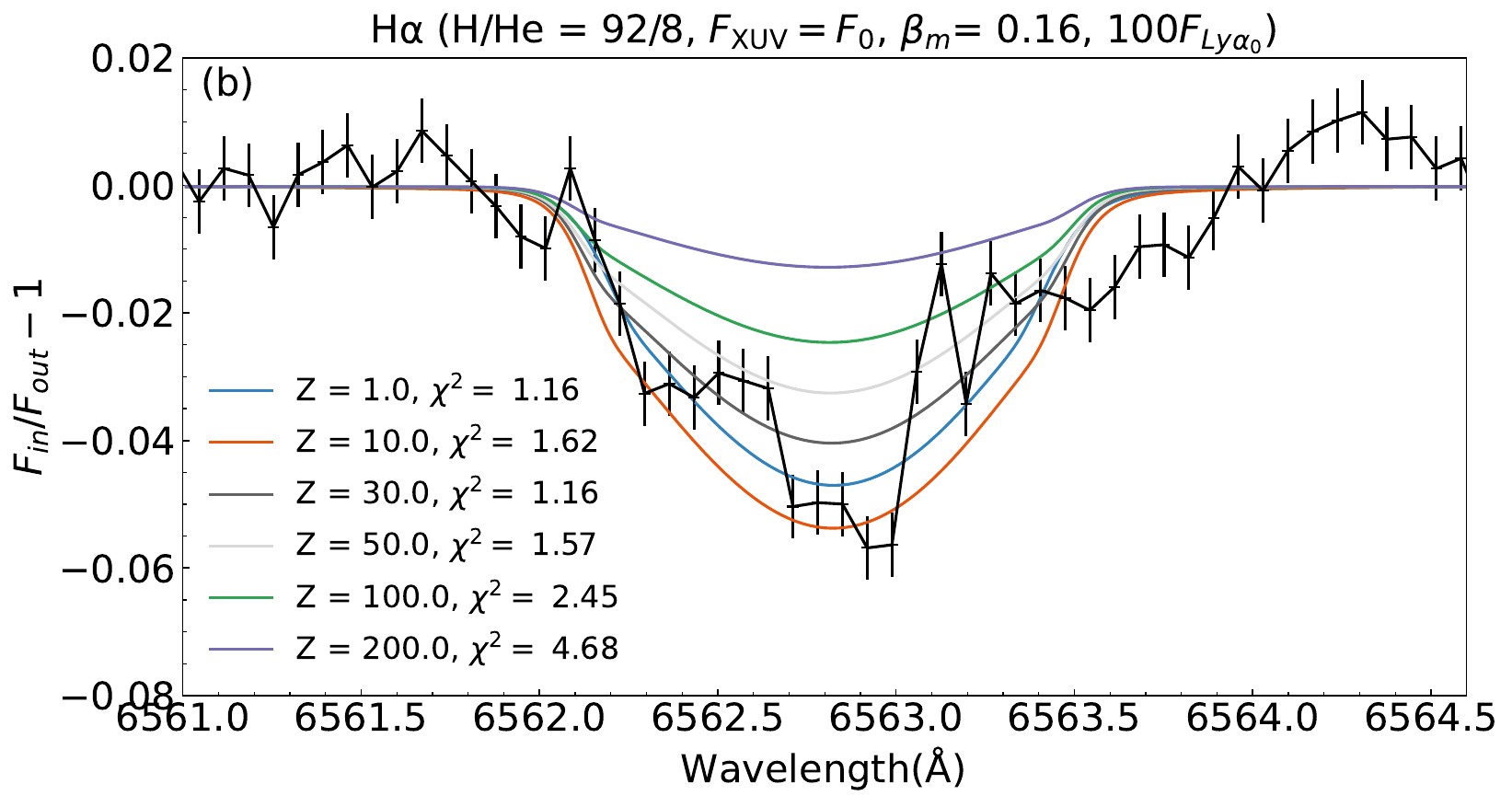}
	\end{minipage}
    \caption{Comparison of  He 10830 and H$\alpha$ transmission spectra with the models of H/He = 92/8, $F_{\rm XUV} = F_0$, and $\beta_m = 0.16$ for metallicities ranging from Z = 1 to Z = 200. The stellar Ly$\alpha$ flux is 100 times the fiducial value, i.e., $F_{\rm Ly\alpha} = 100 F_{\rm Ly\alpha_0}$.}
    \label{fig:model_TS_Z6}
\end{figure}

\begin{figure*}
    \centering
    \subfigure{\includegraphics[width=0.49\textwidth]{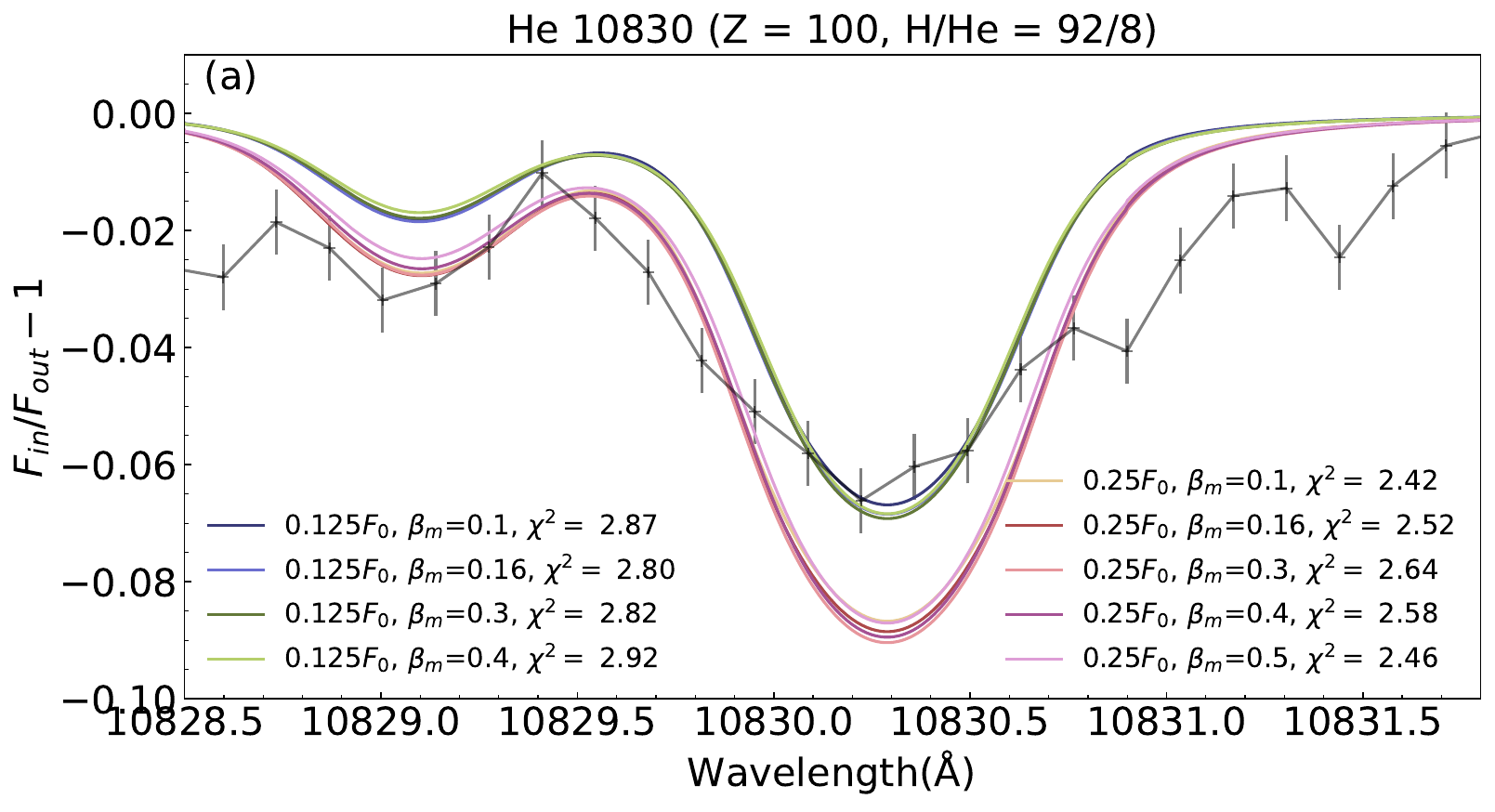}} 
    \subfigure{\includegraphics[width=0.49\textwidth]{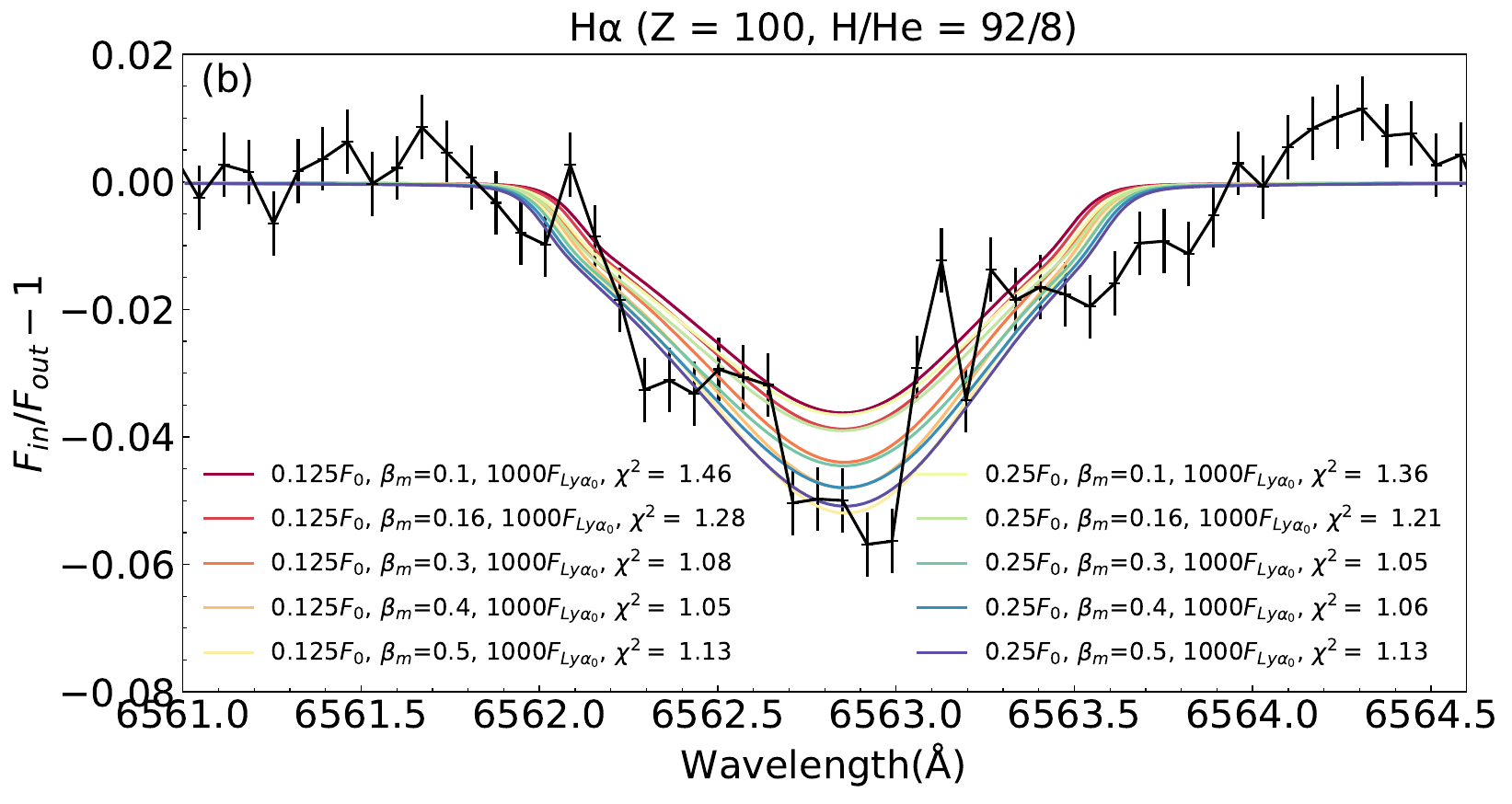}}
    \\ 
    \centering
    \subfigure{\includegraphics[width=0.49\textwidth]{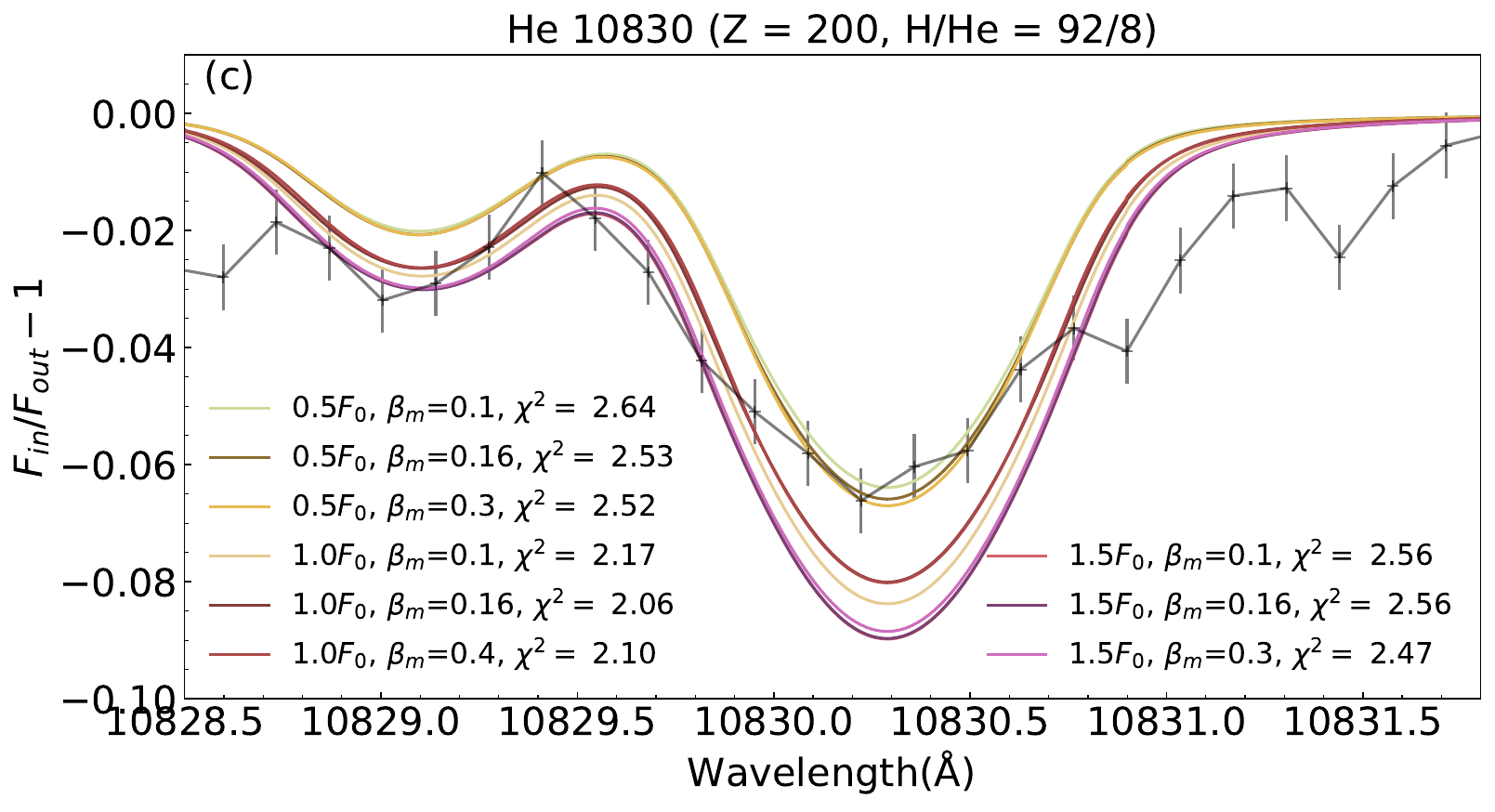}}
    \subfigure{\includegraphics[width=0.49\textwidth]{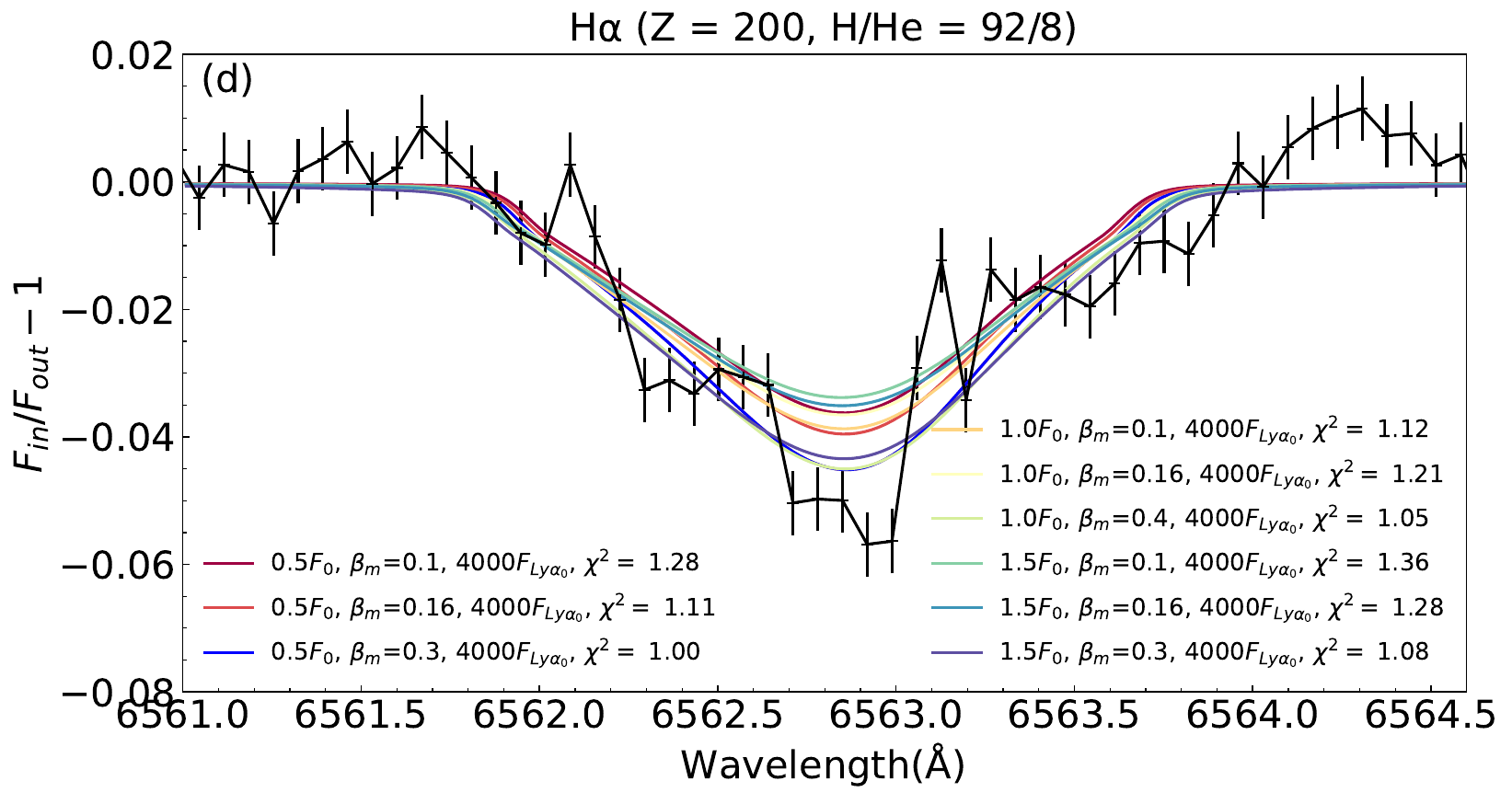}}
        \\ 
    \centering
    \subfigure{\includegraphics[width=0.49\textwidth]{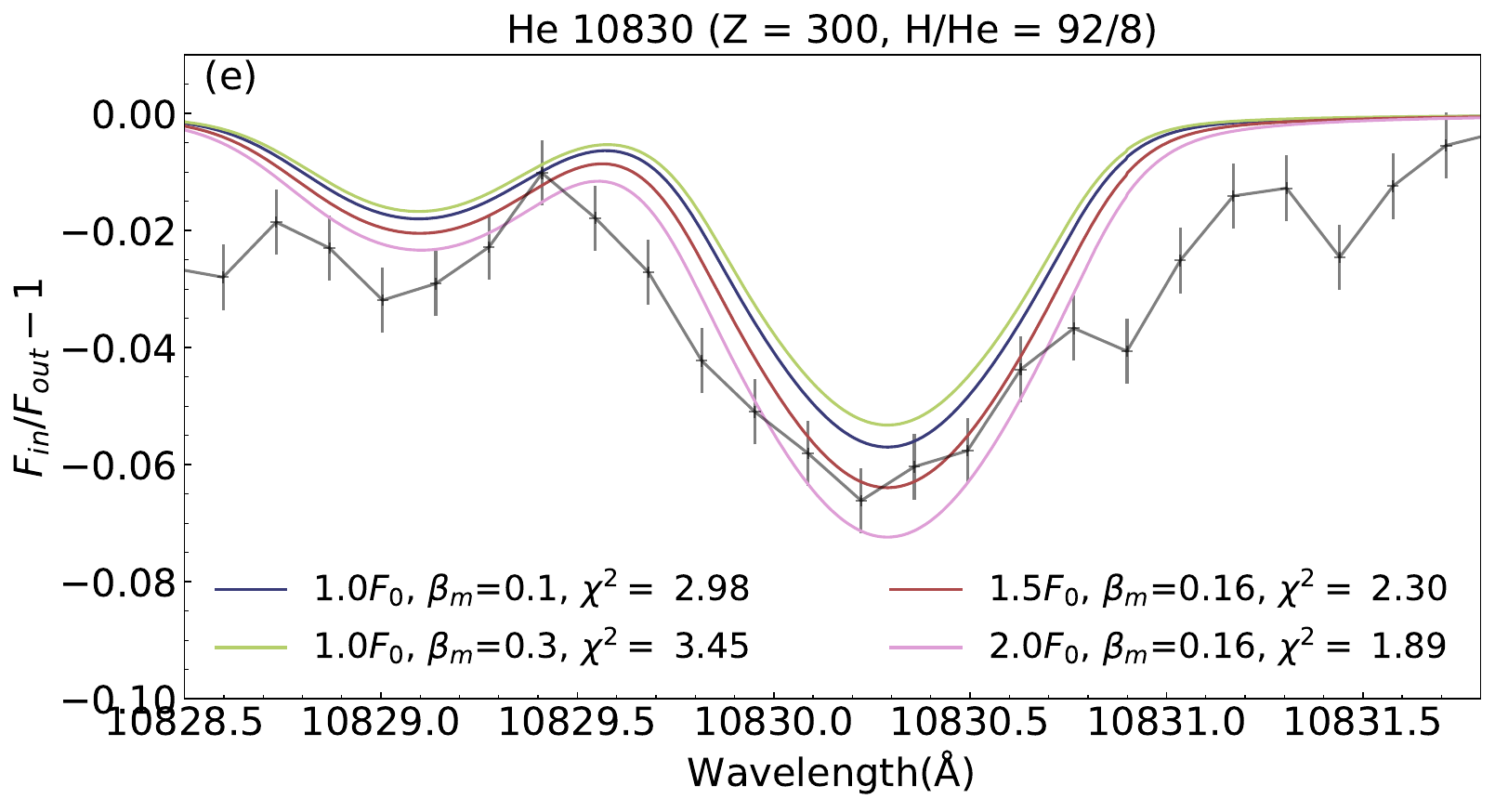}}
    \subfigure{\includegraphics[width=0.49\textwidth]{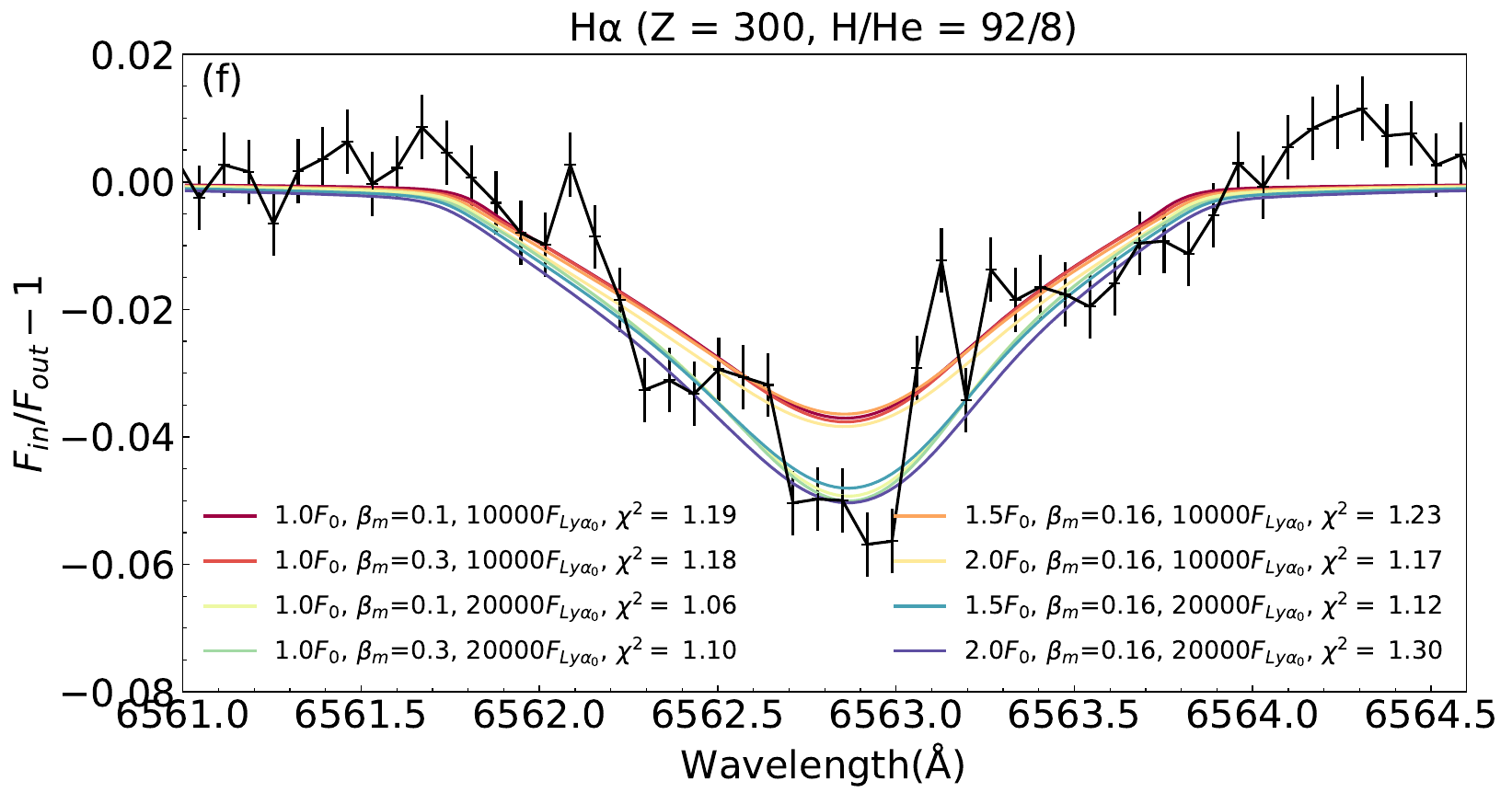}}
    \caption{Comparisons of the He 10830 and H$\alpha$ transmission spectra with the observations, for the models with Z = 100, 200, and 300. The lines with error bars are the observation data from \cite{2022AA...657A...6C}.} 
    \label{fig:He10830-Halpha-z300}
\end{figure*}

\subsection{Modelling He 10830 and H$\alpha$ transmission spectra with super-solar metallicity (Z > 1)}\label{sec:Result-part1-4}

According to \cite{2020AJ....160...51A}, the lower atmosphere of HAT-P-32b has a metallicity of $log(Z/Z_{\sun}) = 2.41 ^{+0.06}_{-0.07}$ (note this $Z$ is different to the dimensionless Z used in our work), corresponding to 218-300 times the solar metallicity. Is it possible that the upper atmosphere also exists a super-solar metallicity? To answer this question, we investigated if the metals can rise into the upper atmosphere and eventually escape. \cite{1987Icar...69..532H} proposed that heavy species can be dragged by light ones through collisions if the mass of heavy species is smaller than the crossover mass \citep{2023ApJ...953..166X}. Here, we examined the oxygen species in the atmosphere of HAT-P-32b, and found that the mass of oxygen is smaller than its crossover mass. This indicates that oxygen can be dragged into the upper atmosphere. Other heavy species may also escape because the crossover mass is one or two orders of magnitude higher than the mass of oxygen. It's important to note that we did not consider metal cooling in this work; thus, the escaping flux may be overestimated to some degree and so could be the crossover mass. However, an upper atmosphere with a super-solar metallicity but with a solar H/He ratio seems unlikely. In any case, as we show below, from the joint analysis of He 10830 and H$\alpha$ absorptions, a super-solar upper atmosphere can be ruled out.

Increasing the metallicity in the upper atmosphere can cause a significant change to the atmospheric structures and thus to the He 10830 and H$\alpha$ absorption profiles. Figure \ref{fig:atm_cases_z} shows the atmospheric structures for the models with H/He = 92/8, $F_{\rm XUV} = F_0$, and $\beta_m = 0.16$, for metallicities ranging from Z = 1 to Z = 200. Figure \ref{fig:model_TS_Z6} shows the He 10830 and H$\alpha$ transmission spectra of these models.

Figure \ref{fig:atm_cases_z} (a) shows the atmospheric temperature and velocity. The temperature increases by XUV heating, and decreases primarily by the adiabatic expansion. The peak temperature increases with increasing metallicity.
The velocity increases with altitude and become supersonic beyond the sonic points (denoted by the asterisks in the plot). At the outer boundary of the atmosphere, it exceeds about $80 \ \rm km \ s^{-1} $. As the metallicity increases, the velocity decreases, and the sonic points tend to be at higher altitudes. 
Figure \ref{fig:atm_cases_z} (b) shows the densities of H(1s) and H$^+$. 
H(1s) dominates at small radii, while H$^+$ becomes more abundant beyond $r \approx 1.1-1.3R_P$. 
For Z < 10, an increase of metallicity would lead to higher number densities of H(1s) and H$^+$, while the opposite holds true for Z > 10. High metallicity (Z > 10) also tends to create to an ionization front at lower altitude, the altitude where the number densities of H(1s) and H$^+$ are equal.
Figures \ref{fig:atm_cases_z} (c-d) show the densities of He, He$^+$, and metastable H(2$^3$S). The number density of He(2$^3$S) has a similar profile to that of He$^{+}$, indicating a close relationship between the production of He(2$^3$S) and the number density of He$^{+}$. In fact, studies have shown that the recombination of He$^{+}$ is the dominant process of producing He(2$^3$S) \citep{2019ApJ...881..133O,2021A&A...647A.129L,2022AA...657A...6C, 2022ApJ...936..177Y}.
Similar to hydrogen, the number densities of helium species decrease with the increase in metallicity for Z > 10, but the opposite occurs for Z < 10. \cite{2022AJ....163...67Z} also found that the population of metastable helium does not change monotonously with metalliciy.

As a result, the He 10830 and H$\alpha$ absorptions first increase with the metallicity when Z < 10, and then decrease significantly when Z > 10, as can be seen from Figure \ref{fig:model_TS_Z6}. For the model with H/He = 92/8, $F_{\rm XUV} = F_0$, and $\beta_m = 0.16$, solar metallicity can lead to an absorption depth of He 10830 higher than 30\%, and it can be reduced to less than 10\% when Z = 200. Introducing a higher metallicity, as suggested by the transmission spectroscopy of the lower atmosphere \citep{2020AJ....160...51A}, may possibly mitigate the requirement for a high H/He ratio in order to fit the He 10830 absorption, as discussed.

In Figure \ref{fig:He10830-Halpha-z300}, we show the transmission spectra of He 10830 and H$\alpha$ for models with Z = 100, 200, and 300, when keeping the H/He ratio as 92/8. For these models, the reduced $\chi^2$ values are all larger than 1.6 (3$\sigma$), so we showed the results with $\chi^2 < 3$. It can be seen that for Z = 100, only models with $F_{\rm XUV} \leq 0.25F_0$ can give $\chi^2 < 3$ for the He 10830 transmission spectra. The He 10830 absorption changes slightly with the spectral index $\beta_m$. For these models, a stellar Ly$\alpha$ flux as high as 1000 times the fiducial value is required to explain the H$\alpha$ observation. For Z = 200, $F_{\rm XUV}$ is about (0.25- 1.5) $F_0$, and $F_{\rm Ly\alpha}$ should be about $4000 F_{\rm Ly\alpha_0}$; and for Z = 300, $F_{\rm Ly\alpha}$ increases to 10,000-20,000 times the fiducial value!
As shown in the Section 3.1, such high $F_{\rm Ly\alpha}$ is unlikely. Therefore, while a super-solar metallicity (100 times higher than the solar value) could occur in the lower atmosphere of HAT-P-32b, a nearly solar metallicity can be present in the upper atmosphere.

Therefore, while a super-solar metallicity (100 times higher than the solar value) could occur in the lower atmosphere of HAT-P-32b, a nearly solar metallicity may be present in the upper atmosphere. On the one hand, it's possible that light species (hydrogen and helium), and even metals in the form of atoms, ions, or some molecules, rise into the upper atmosphere. However, the escaping metals may only constitude a small proportion of the total metals. As suggested by \cite{2020AJ....160...51A}, the high metallicity of HAT-P-32b in their study is probably indicative of a thick cloud deck or haze, where metals exist in the form of submicron-sized particles, such as magnesium silicate \citep{2017AN....338..773M}, which cannot rise into the upper atmosphere.   On the other hand, the escaping hydrogen and helium also contributes to a reduction in the mass fraction of metals in the upper atmosphere.

\subsection{Best-fit models}\label{sec:Result-part1-1}

\begin{figure}
	\begin{minipage}[t]{0.5\textwidth}
		\centering
		\includegraphics[width=\textwidth]{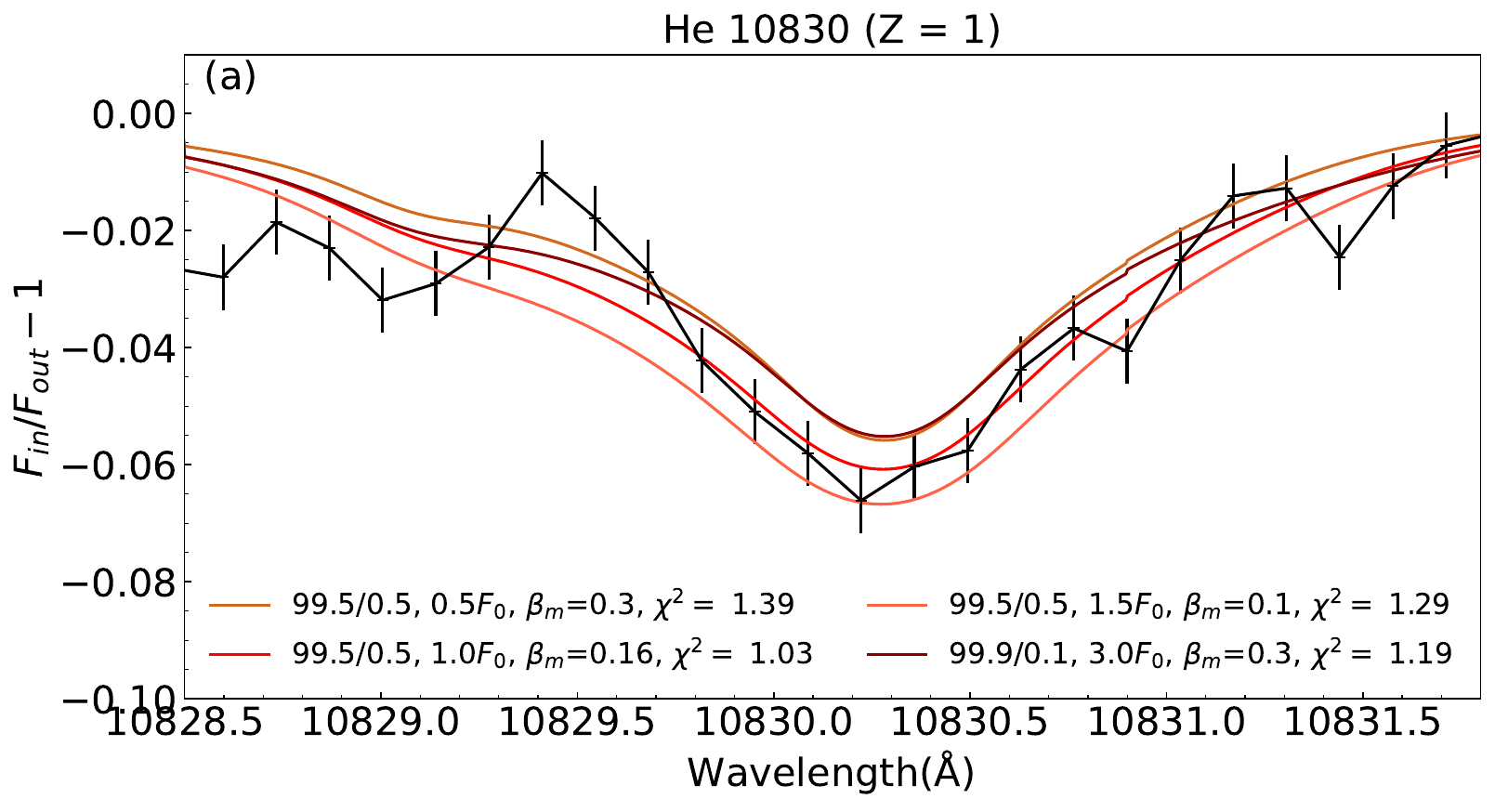}
	\end{minipage}
	\begin{minipage}[t]{0.5\textwidth}
		\centering
		\includegraphics[width=\textwidth]{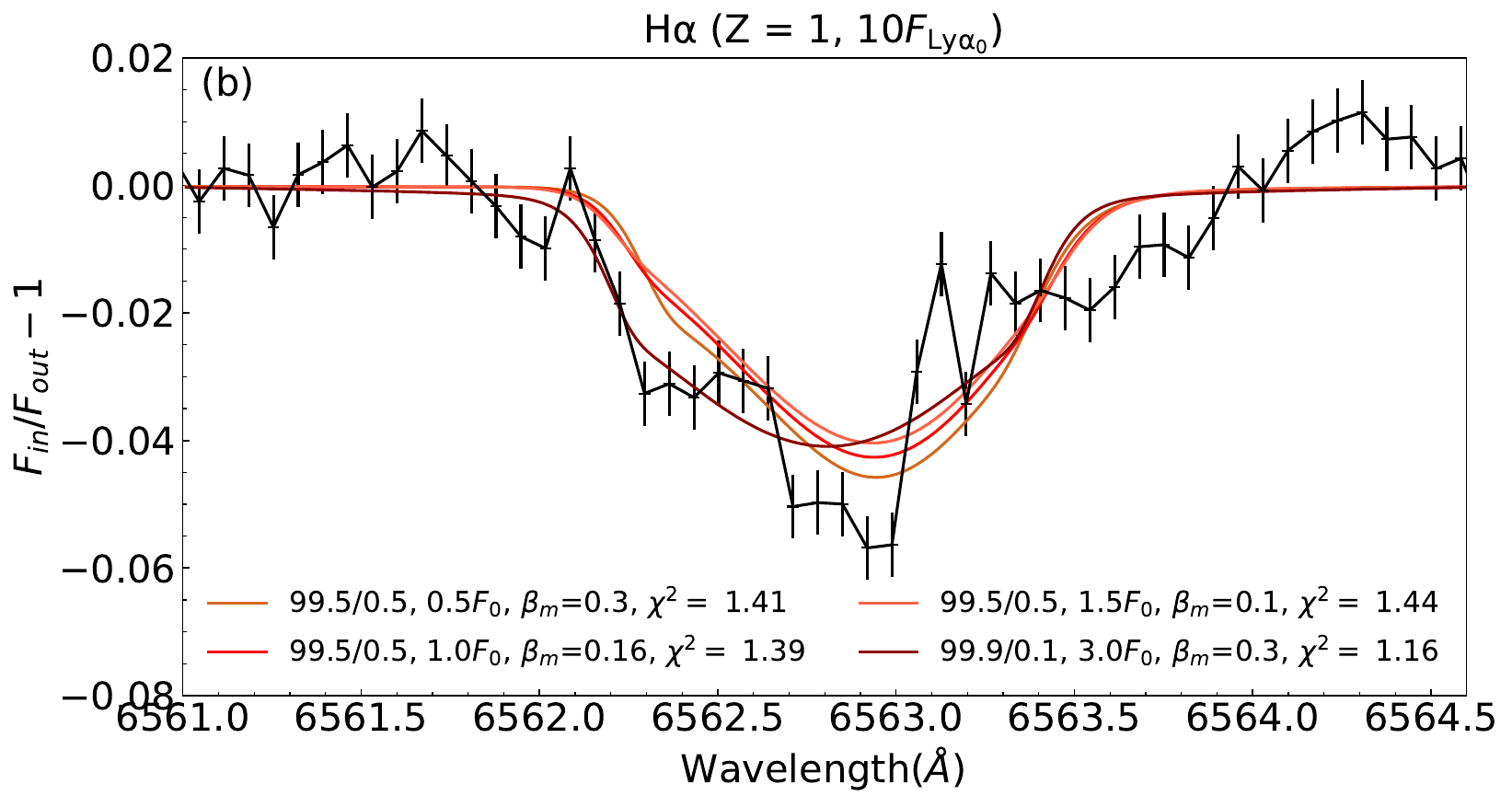}
	\end{minipage}
    \caption{The model transmission spectra that fit both the He 10830 and H$\alpha$ observations. The lines with error bars are the observation data from \cite{2022AA...657A...6C}.} 
    \label{fig:He10830-Halpha-bestfit}
\end{figure}

In summary, we found four best-fit models that reproduce both observations simultaneously: (1) Z = 1, H/He =99.5/0.5, $F_{\rm XUV}= 0.5 F_0$, $\beta_m = 0.3$, and the mass-loss rate $\dot{M} = 1.0 \times 10^{13} \rm  g\ s^{-1}$; (2) Z = 1, H/He =99.5/0.5, $F_{\rm XUV}= 1.0 F_0$, $\beta_m = 0.16$, and $\dot{M} = 1.5 \times 10^{13} \rm g\ s^{-1}$; (3) Z = 1, H/He =99.5/0.5, $F_{\rm XUV}= 1.5 F_0$,  $\beta_m = 0.1$, and $\dot{M} = 1.8 \times 10^{13} \rm g\ s^{-1}$; (4) Z = 1, H/He =99.9/0.1, $F_{\rm XUV}= 3.0 F_0$, $\beta_m = 0.3$, and $\dot{M} = 3.1 \times 10^{13} \rm g\ s^{-1}$. The He 10830 and H$\alpha$ transmission spectra of these models are shown in Figure \ref{fig:He10830-Halpha-bestfit}. The resulting mass-loss rate of about, approximately (1.0 $\sim$ 3.1) $\times 10^{13}$g s$^{-1}$, $\ $is consistent with those from the energy-limited approach applying a very low heating efficiency, and with those of \cite{2022AA...657A...6C} and \cite{2023A&A...673A.140L}. The H/He ratio of $\geq$ 99.5/0.5 is also consistent with that of \cite{2023A&A...673A.140L}, who conclude that the H/He ratio in the upper atmosphere of HAT-P-32b is (99.0/1.0)$^{+0.5}_{-1.0}$. The H/He ratio of approximately 99.5/0.5 is significantly higher than the solar value (92/8), indicating a substantial reduction in the helium abundance in the upper atmosphere. 
There exists a degeneracy in the combination of the XUV flux and spectral index $\beta_m$. Our fiducial model assumes $F_{\rm XUV}= 1.0 F_0$ and $\beta_m = 0.16$, the same values as adopted in \cite{2022AA...657A...6C} and \cite{2023A&A...673A.140L}.
Other combinations, such as $F_{\rm XUV}= 0.5 F_0$ and $\beta_m = 0.3$,  $F_{\rm XUV}= 1.5 F_0$ and $\beta_m = 0.1$, $F_{\rm XUV}= 3.0 F_0$ and $\beta_m = 0.3$, also give reasonable results. The XUV flux was obtained using the XMM-Newton data observed on 30 August 2019, but the H$\alpha$ and He 10830 transmission data were observed on 1 September and 9 December 2018, respectively. 
The XUV ﬂux might have been lower in 2018 when H$\alpha$ and He 10830 lines were observed. At the same time, it is also possible that the X-ray fraction could be 30\% of the XUV ﬂux, higher than in the ﬁducial model, because the stellar activity can vary over time in such an active host star.
It is also possible that both the XUV flux $F_{\rm XUV}$ and the spectral index $\beta_m$ exceed the fiducial values, as indicated by the fourth best-fit model, due to a high stellar activity.

\subsection{The atmospheric structures of the best-fit models}\label{sec:Result-part1-1}

\begin{figure*}
    \centering
    \subfigure {\includegraphics[width=0.49\textwidth]{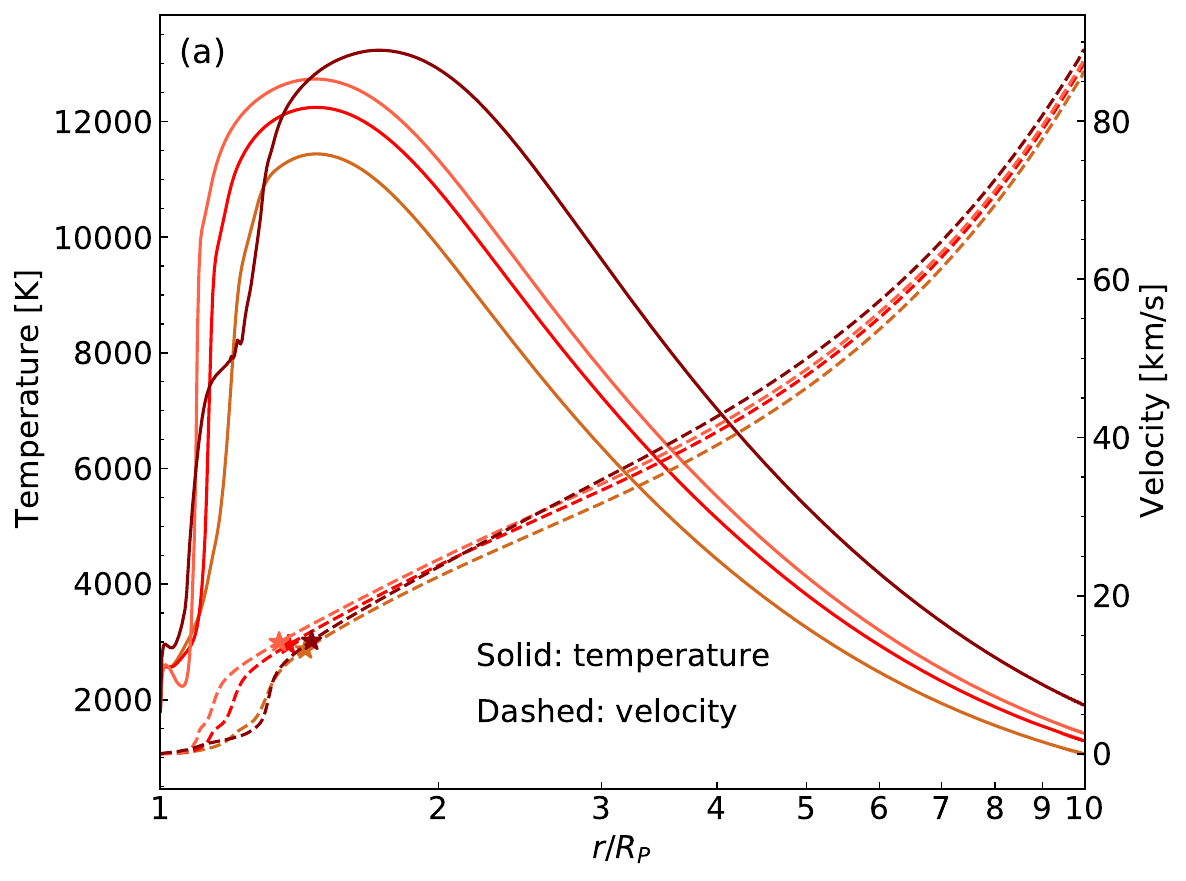}}
    \subfigure {\includegraphics[width=0.5\textwidth]{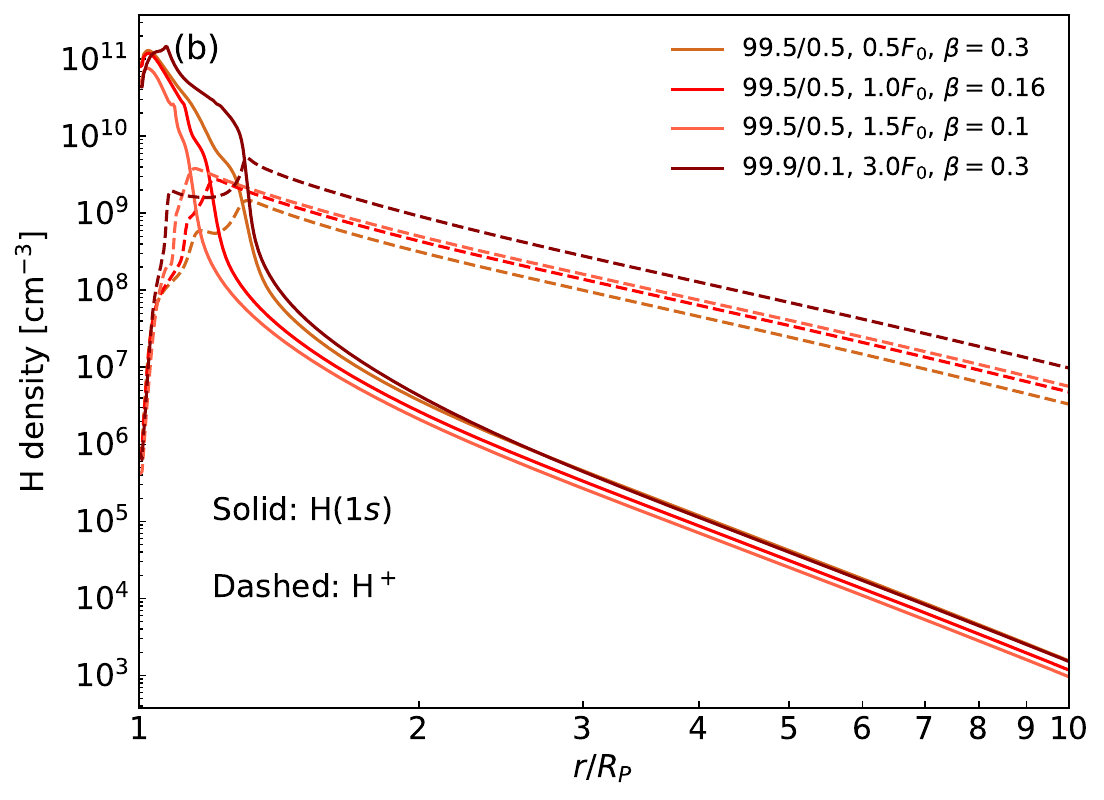}}
    \\ 
    \centering
    \subfigure{\includegraphics[width=0.49\textwidth]{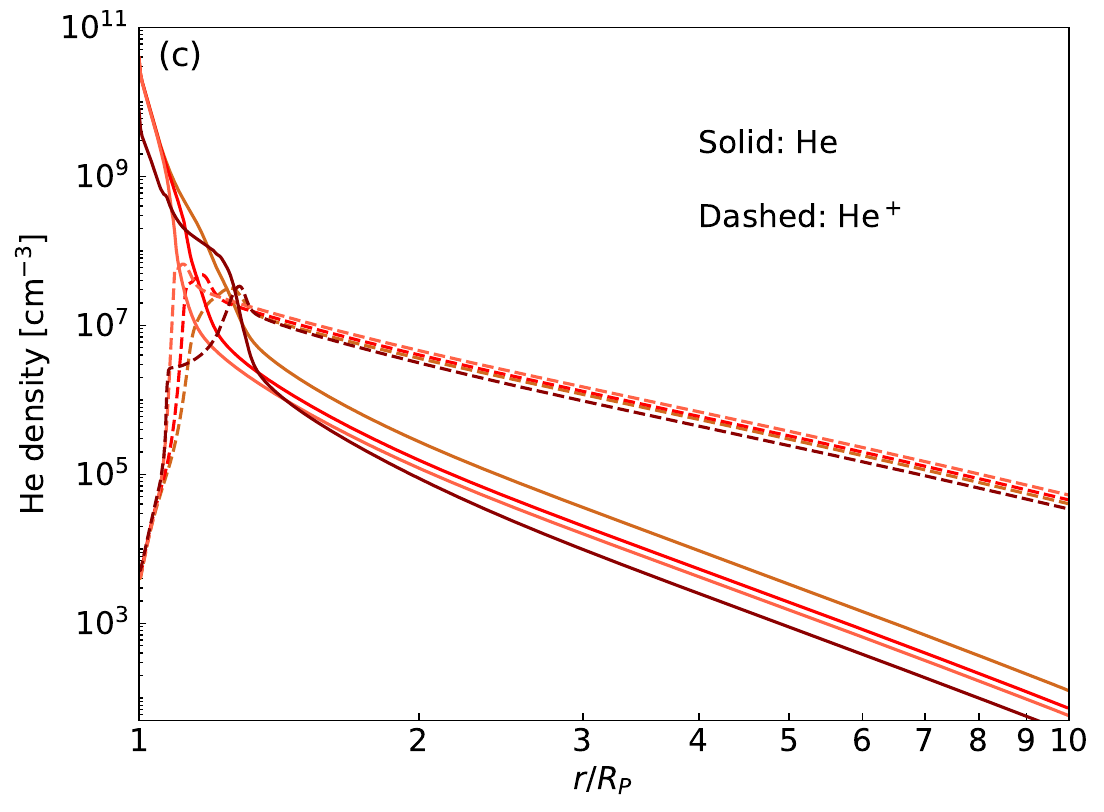}}
    \subfigure {\includegraphics[width=0.5\textwidth]{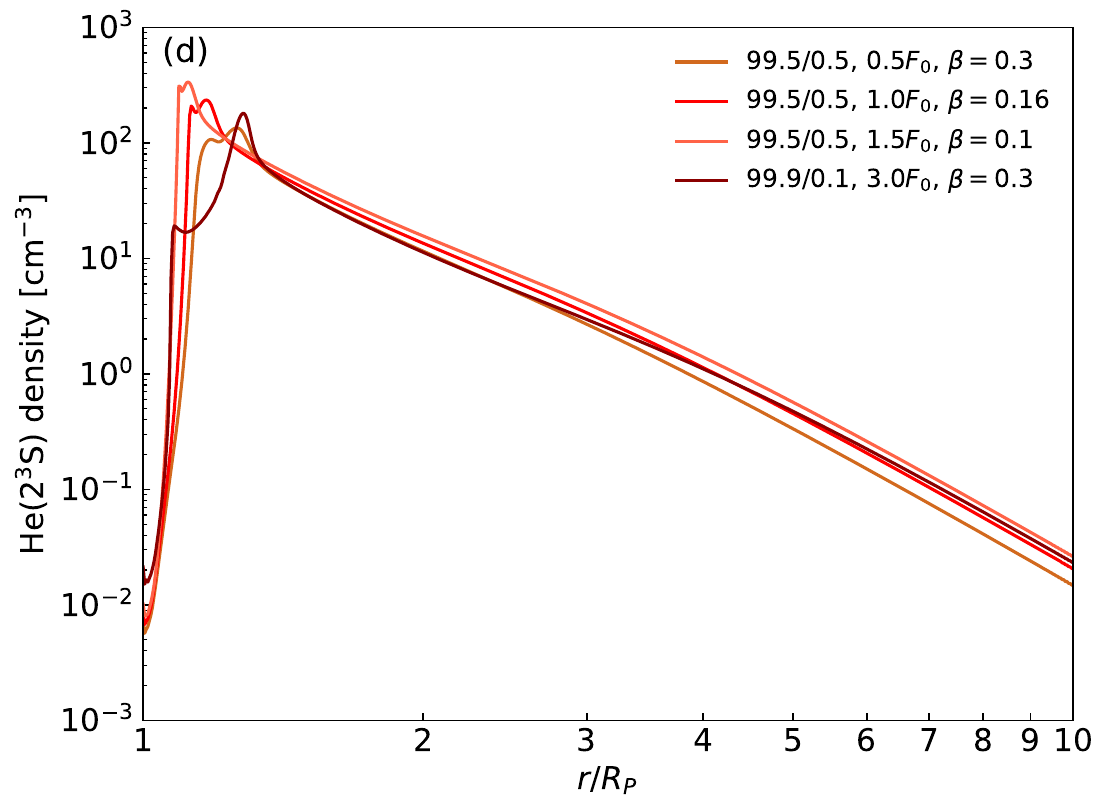}}
    \caption{The atmospheric structure of HAT-P-32b for the best-fit models. (a) Temperature (denoted in solid lines) and velocity (dashed lines). The asterisks on the velocity lines represent the sonic points. (b) Number densities of hydrogen atoms H(1s) and ions H$^+$. (c) Number densities of helium atoms He and ions He$^+$. (d) Number densities of helium atoms in the metastable state, He(2$^3$S).} 
    \label{fig:atm_cases_bestfit}
\end{figure*}

In this section, we present the atmospheric structures and Ly$\alpha$ radiation intensity within the atmosphere for the best-fit models. 
Figure \ref{fig:atm_cases_bestfit} shows the atmospheric temperature, velocity, and number densities of the hydrogen and helium species. They exhibit similar trends to those in Figure \ref{fig:atm_cases_z}. However, it's noteworthy that the temperatures for the best-fit models fall within the range of 11,400K to 13,200K, consistent with the results reported by \cite{2023A&A...673A.140L}. Additionally, the velocity near 2.4$R_P$ is also similar to that of \cite{2022AA...657A...6C} and \cite{2019ApJ...884L..43G}. 

\begin{figure*}
    \centering
    \subfigure{\includegraphics[width=0.49\textwidth]{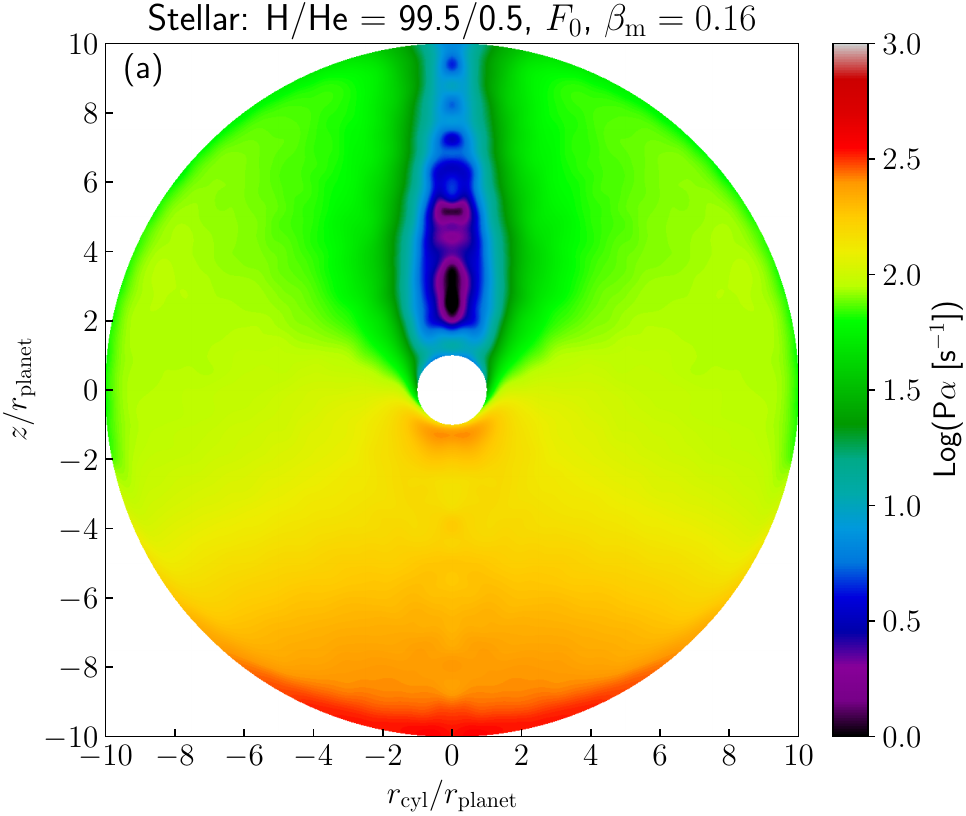}} 
    \subfigure{\includegraphics[width=0.49\textwidth]{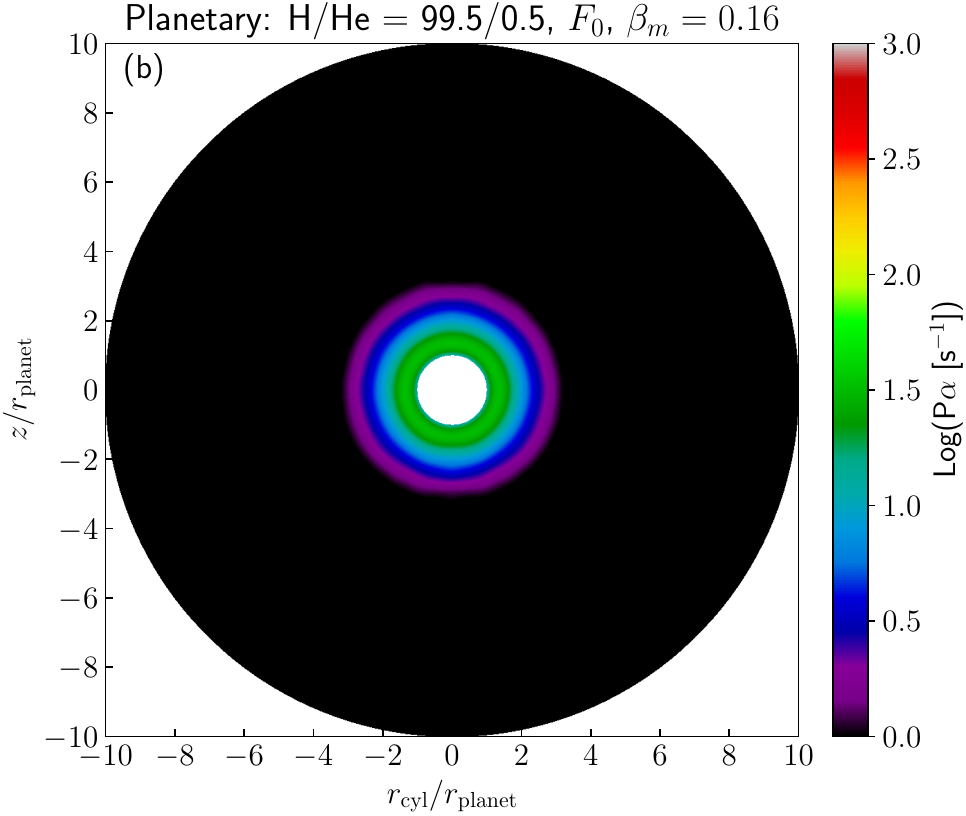}}
    \\ 
    \centering
    \subfigure{\includegraphics[width=0.49\textwidth]{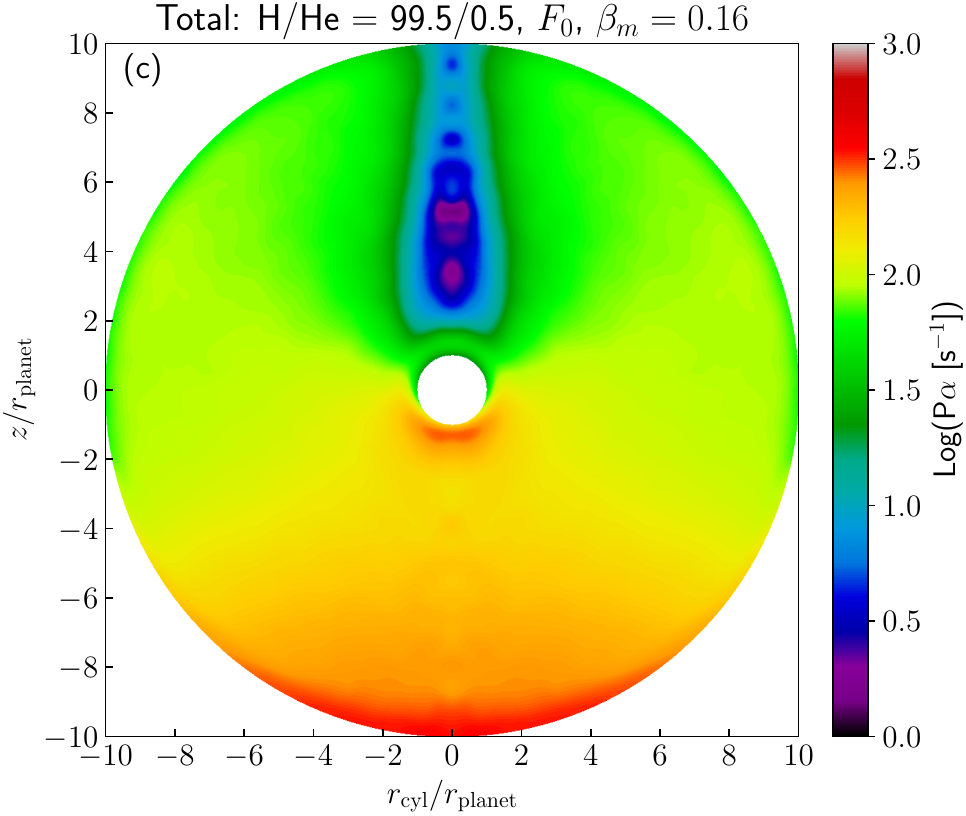}}
    \subfigure{\includegraphics[width=0.49\textwidth]{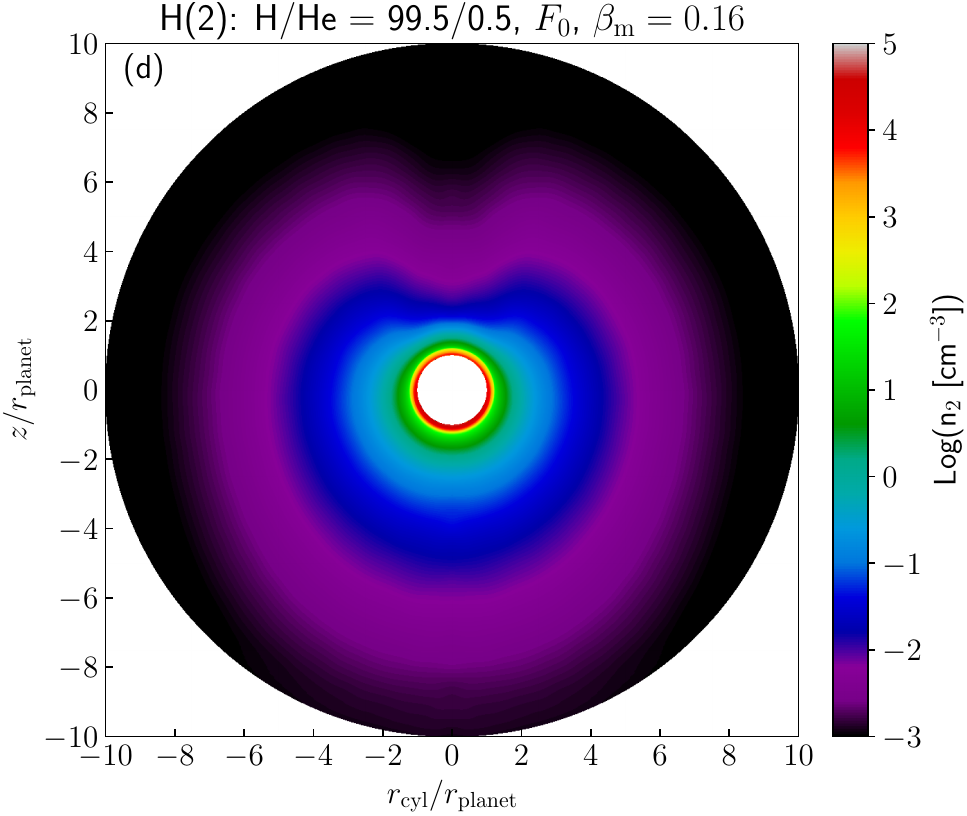}}
    \caption{The scattering rate $P\alpha$ and H(2) number density distributions. (a-c) $P\alpha$ obtained for the stellar, planetary Ly$\alpha$, and both respectively. (d) H(2) number density. Note that all the colorbars are given in log scale.} 
    \label{fig:atm_pa_H2_spherical}
\end{figure*}

Using the atmospheric structures of H(1s) and H$^+$, we performed the Ly$\alpha$ radiative transfer simulation, obtaining the Ly$\alpha$ radiation intensity within the atmosphere and the H(2) population responsible for the H$\alpha$ absorption. 
Figure \ref{fig:atm_pa_H2_spherical} shows the scattering rate $P\alpha$ and H(2) number density distributions in the cylindrical coordinates of the atmosphere obtained from the best-fit model. 
The scattering rate $P\alpha = B_{12}\bar{J}_{\rm Ly\alpha} \ [\rm s^{-1}]$ is the number of scatterings per second experienced by a hydrogen atom, where $\bar{J}_{\rm Ly\alpha}$ and $B_{12}$ are the Ly$\alpha$ radiation intensity averaged over the solid angle and the Einstein absorption $B$ coefficient, respectively. Higher scattering numbers per second per H atom are related to a higher Ly$\alpha$ intensity, thus cause larger H(2) populations. 
For the sake of brevity, we only show the results of one representative model with H/He = 99.5/.5, $F_{\rm XUV} = F_0$, and $\beta_m = 0.16$. Similar results are found for the other three best-fit models.
In Figure \ref{fig:atm_pa_H2_spherical}, the white disks denote the planet and the surroundings are the atmosphere. The $z-$axis connects the centers of the star and the planet. In panels (a) and (c), the stellar Ly$\alpha$ photons are incident from the bottom. Figure \ref{fig:atm_pa_H2_spherical}(a) shows $P\alpha$ calculated for the stellar Ly$\alpha$, which has a cylindrical symmetry. One can see that the $P\alpha$ in the dayside is much higher than that in the nightside for the stellar Ly$\alpha$ case. The result is similar to what we found in Paper I. $P\alpha$ calculated for the planetary Ly$\alpha$ shown in Figure \ref{fig:atm_pa_H2_spherical}(b), however, is spherically symmetric. Comparing Figures \ref{fig:atm_pa_H2_spherical}(a) and (b), we find that in most regions (except for a small part in the shadow of the planet), $P\alpha$ by the planetary Ly$\alpha$ is much lower than by the stellar one and is negligible. A similar trend was found in Paper I. However, this is different from the results of HD 189733b and WASP-121b reported by \cite{2017ApJ...851..150H,2023ApJ...951..123H}. In their work, $\bar{J}_{\rm Ly\alpha}$ resulting from recombination inside the planetary atmosphere can exceed that of the external stellar Ly$\alpha$ in a certain altitude range for HD 189733b; and $\bar{J}_{\rm Ly\alpha}$ from collisional excitation can become dominant in some regions for WASP-121b. The difference between our results and theirs could be due to the differences in the planetary parameters and the atmospheric geometries (spherical in our study vs. plane-parallel in theirs).

\section{Discussion}\label{sec: discuss}

\subsection{The effect of Ly$\alpha$ cooling and heating}

When a Ly$\alpha$ photon is emitted by an atom, the atom loses energy unless the photon is reabsorbed. During the resonant scattering processes of Ly$\alpha$ photons, energy is conserved in the rest frame of the atom if the recoils of hydrogen atoms are negligible. Consequently, the gas cools down as it emits Ly$\alpha$ photons.
Ly$\alpha$ photons created in the atmosphere have three destinations. First, they can escape the atmosphere after many resonance scatterings, which plays an important role in cooling the atmosphere. Second, they can be absorbed by dust grains and molecular hydrogen \citep{1990ApJ...350..216N}. Third, they can be absorbed by the planetary surface, which can be regarded as cooling of the atmosphere. 
In our simulations, there is no dust but molecular hydrogen H$_2$ at the lower boundary of the upper atmosphere. Molecular hydrogen will be rapidly dissociated or ionized by stellar XUV radiation.
However, the resonance-like absorption of Ly$\alpha$ by molecular hydrogen is not considered in our simulation. Therefore, the Ly$\alpha$ photons escape freely from the atmosphere in our simulations.

In this work, the Ly$\alpha$ cooling due to collisional excitation is considered in the hydrodynamic simulations. The cooling rate is calculated using $\Lambda_{\rm Ly\alpha} = -7.5\times10^{-19}n_en_{1s}e^{-118348K/T} \rm erg \ cm^{-3} \ s^{-1}$ \citep{1981MNRAS.197..553B,2009ApJ...693...23M}, where $n_e$ and $n_{1s}$ are the number densities of electrons and neutral hydrogen atoms, respectively. Ly$\alpha$ photons can also be generated through the recombination of H$^+$. Ignoring this process may result in some underestimation of the cooling.

However, the recoil of hydrogen atoms due to resonance scattering transfers a portion of Ly$\alpha$ energy to the hydrogen atoms, eventually heating the hydrogen gas. Depending on the scattering angle, an atom's recoil can result in either energy gain or loss, but when averaged over the entire scattering angle, it ultimately leads to net heating.
The heating induced by Ly$\alpha$ scattering has been studied in cosmology \citep{1997ApJ...475..429M,2004ApJ...602....1C}. However, there is some debate regarding the significance of the heating rate. \cite{1997ApJ...475..429M} first pointed out the heating from Ly$\alpha$ scattering and demonstrated its potential criticality. In contrast, \cite{2004ApJ...602....1C} argued that the heating induced by Ly$\alpha$ scattering is negligible. It is unclear how significant the heating by recoil of hydrogen atoms caused by resonance scattering would be in the planetary atmosphere. However, the heating rate will be likely to be significantly smaller than the cooling rate induced by collisional excitation and recombination. This is because the heating by recoil is an average over the scattering angle. It was also found that the imprint of the recoil effect on the Ly$\alpha$ spectral profile near the line center is not clearly appreciable in a high-temperature ($T\sim 10^4\ K$ ) gas in the context of Wouthuysen-Field effect, which is closely associated with the Ly$\alpha$ heating \citep{2020ApJS..250....9S}.

\subsection{The effects of turbulence and limb darkening}

In this work, the Voigt profile is used when calculating the absorption profiles of He 10830 and H$\alpha$. The hydrodynamic bulk velocity and the thermal velocity contribute to the line broadening. In addition to these broadening components, there can be turbulence in the atmosphere.
The turbulence effect can be described by adopting an effective Doppler width $\Delta\nu_D = \frac{\nu_0}{c}(\frac{2kT}{m}+v_{turb}^2)^{1/2}$ in place of the thermal Doppler width, where $v_{turb}$ is the turbulence velocity \citep{1986rpa..book.....R,2020ApJS..250....9S}. 
The line broadening by turbulence has been included when calculating the transmission spectra of exoplanetary atmosphere in some works \citep{2018A&A...620A..97S,2020A&A...636A..13L,2022AA...657A...6C}. These authors assumed that the turbulence velocity $v_{turb} = \sqrt{5kT/3m}$, which corresponds to the speed of sound ($c_s$) of a monatomic gas, and thus to the transonic case with a Mach number of 1.

\begin{figure}
	\begin{minipage}[t]{0.5\textwidth}
		\centering
		\includegraphics[width=\textwidth]{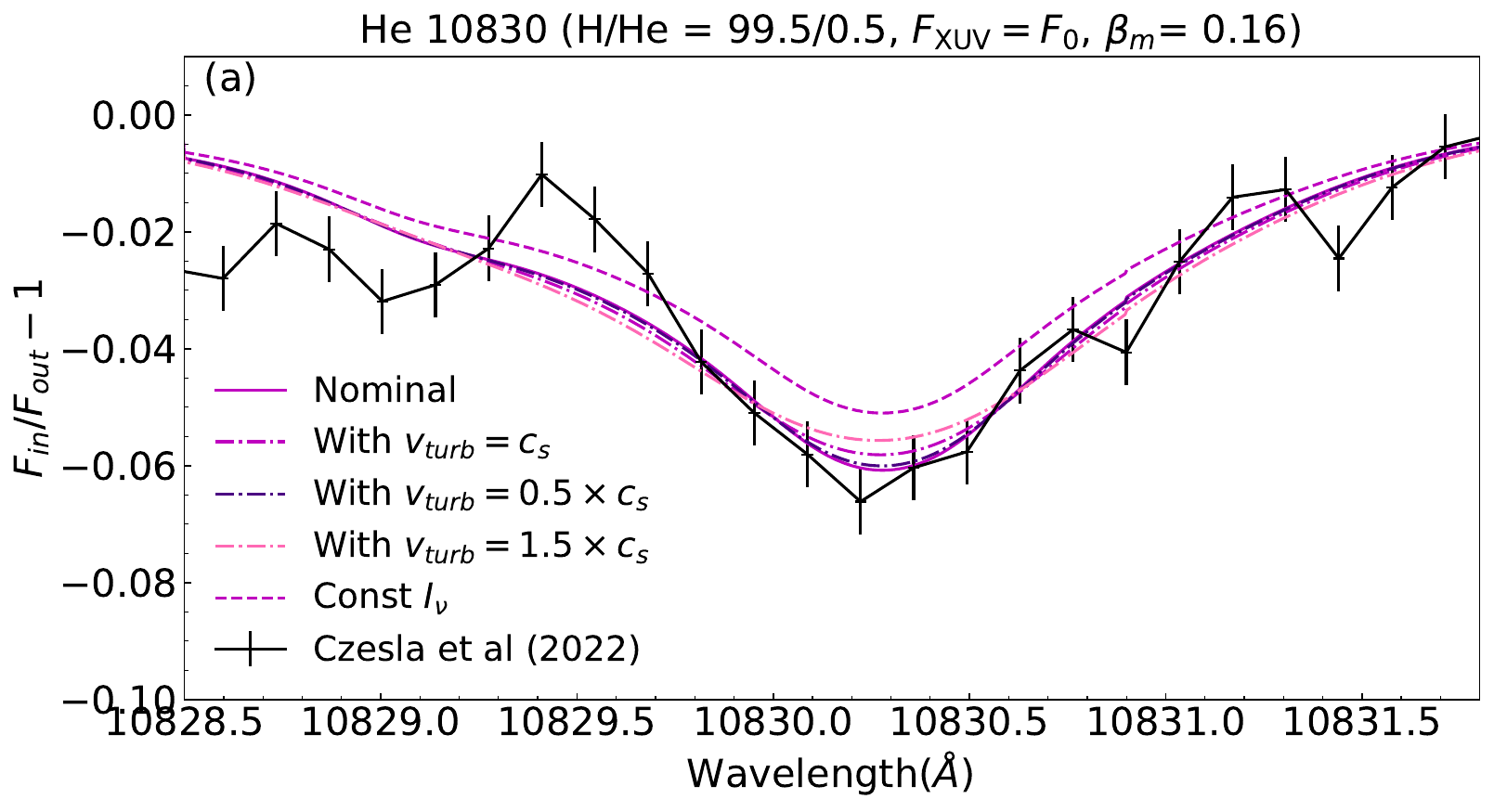}
	\end{minipage}
	\begin{minipage}[t]{0.5\textwidth}
		\centering
		\includegraphics[width=\textwidth]{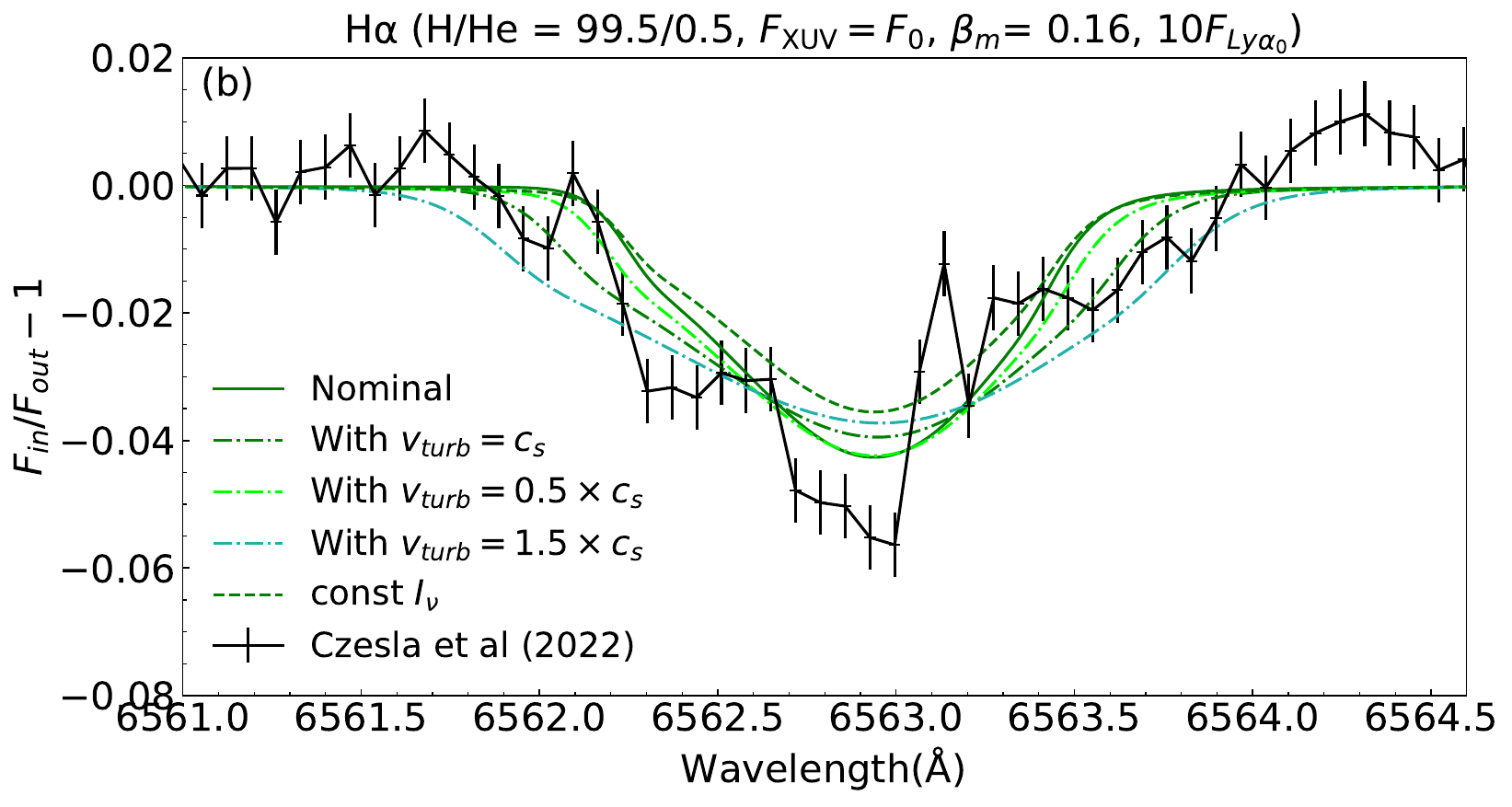}
	\end{minipage}
    \caption{Transmission spectra of He 10830 and H$\alpha$ lines for the models with different turbulence velocities and limb-darkening laws. The solid, dash-dotted, and dashed lines represent the cases of the nominal models, the models including the turbulence effect with a velocity of 0.5 to 1.5 times the speed of sound $c_s$, and the model with a constant stellar surface brightness ($I_{\nu} = constant$). The gray lines with error bars are the observation data from \cite{2022AA...657A...6C}.} 
     \label{fig:turbu_effects}
\end{figure}

To evaluate the effects of turbulence in line broadening, we added $v_{turb}$ in the Doppler width assuming the transonic case, and compared the resulting transmission spectra with our nominal models (without turbulence). Figure \ref{fig:turbu_effects} shows the transmission spectra of He 10830 and H$\alpha$ for the models of H/He = 99.5, $F_{\rm XUV} = 1 F_0$, and $\beta_m = 0.16$, when the turbulence velocity varies. The solid and dash-dotted lines represent the nominal models and those with turbulence, respectively. It is found that for He 10830, including the turbulence effect decreases the absorption depth slightly at the line center, but doesn't significantly alter the absorption at the line wings. Similarly, the turbulence leads to a lower H$\alpha$ absorption at the line center, but a deeper absorption at the line wings, resulting in a broader line profile.
In these cases, one would expect a slightly higher $F_{\rm XUV}$ and  $\beta_m$ to fit both the absorption lines. 
However, the turbulence velocity might be different from $v_{turb} = \sqrt{5kT/3m} \ (or \ c_s)$ in a real case, and its effect in the line broadening will be altered accordingly. To demonstrate its effect, we examined the H$\alpha$ and He 10830 transmission spectra of two cases, where $v_{turb} = 0.5 \times c_s$ and $v_{turb} = 1.5 \times c_s$. It can be seen from Figure \ref{fig:turbu_effects} that a larger turbulence velocity tends to lead to a broader absorption line profile.
In the diffuse interstellar medium (ISM), turbulence velocity exhibits an anti-correlation with gas temperature, and the Mach number is typically below 1 at temperatures in the range of a few thousand K or higher (e.g., \citealt{2004ApJ...613.1004R}). Therefore, while it remains uncertain whether the same trend observed in the diffuse ISM can be applied to the atmosphere, the turbulence effect is likely less significant than the one explored in this paper.

Besides turbulence, there are other processes, such as stellar wind and planetary magnetic field, that can affect the line shape \citep{2022ApJ...926..226M,2023arXiv231018486S}. For instance, \cite{2023A&A...673A.140L} estimated the potential effects of stellar winds and found no significant effect on the derived range of temperature and mass-loss rate for HAT-P-32b. A detailed study of the stellar wind and planetary magnetic field requires a comprehensive 3D model, by which we defer that to a future work.

The surface brightness of the stellar disk changes with the limb angle \citep{2015A&A...582A..51C, 2015A&A...574A..94Y}.
We find that adopting different limb-darkening laws results
in variations in the absorption line profiles. In \cite{2022ApJ...936..177Y}, we showed that a constant surface brightness or an isotropic intensity ($I_{\nu} = constant$) gives a lower absorption in both the H$\alpha$ and He 10830 lines than the Eddington limb-darkening law. In this work, we also obtain the same result. The dashed lines in Figure \ref{fig:turbu_effects} show the transmission spectra of the models with a constant stellar surface brightness. The parameters derived in this work may vary slightly if a different limb-darkening law is assumed, but the main results will remain robust.

\subsection{The escape of helium}

In addition to the helium detection of HAT-P-32b by \cite{2022AA...657A...6C}, an escaping helium atmosphere of this planet was also detected by \cite{2023SciA....9F8736Z}. The latter reported giant tidal tails of helium spanning a projected length over 53 times the planet's radius. 
Using a nearly isothermal 3D hydrodynamic model based on Athena ++, \cite{2023SciA....9F8736Z} derived a mass-loss rate of 1.07$\times 10^{12} \rm g \ s^{-1}$, approximately an order of magnitude lower than our result and those of \cite{2022AA...657A...6C} and \cite{2023A&A...673A.140L}.
\cite{2023SciA....9F8736Z} obtained the mass-loss rate by assuming an outflow temperature of about 5750 K and a solar H/He ratio. 
In their analysis of He triplet absorption, \cite{2022AA...657A...6C} and \cite{2023A&A...673A.140L} derived, for a similar temperature and H/He to those of \cite{2023SciA....9F8736Z} (T = 6000 K and an H/He of 90/10), a mass-loss rate slightly lower than $10^{12} \rm g \ s^{-1}$. However, accounting for the 1.5-times-higher absorption measured by \cite{2023SciA....9F8736Z}, we estimate 1.5 times that mass-loss rate ($10^{12} \rm g \ s^{-1}$), in agreement with their result.

As of now, there are several exoplanets with detected escaping helium. On the one hand, models have shown that explaining many of the He 10830 absorption signals may require a higher H/He ratio (or lower helium abundance) than that of the Sun. However, there is still a lack of clear explanations for these elevated H/He ratios. One possible explanation could be that helium, being heavier than hydrogen, is less likely to escape. Therefore, at higher altitudes, the concentration of helium decreases.
Using a multifluid hydrodynamic model, \citet{2023ApJ...953..166X} studied the helium fractionation in the atmosphere of HD 209458b. They found that helium atoms are hard to escape compared to hydrogen, and thus H/He varies with altitude. The low helium abundance could thus be explained by the fractionation mechanism. In our work, the mass of helium is smaller than its crossover mass, which means there is a possibility that helium can be dragged by hydrogen into the upper atmosphere of HAT-P-32b. However, the metal-rich lower atmosphere indicate the portion of hydrogen and helium is reduced in comparison to that of the Sun. 
A high H/He ratio in this planet may indicate a lack of helium in the lower atmosphere in the current stage, or even during the planetary formation. The low helium abundance in the atmosphere implies the necessity of further studies of the helium escape.

On the other hand, one may ask what kind of exoplanets tend to exhibit excess He 10830 absorption. \cite{2019ApJ...881..133O} suggested that close-in exoplanets orbiting late-type stars, especially K-type stars, are more promising to exhibit He 10830 absorption signals. \cite{2018Sci...362.1388N} showed the 10830 excess absorption as a function of the XUV$_{\rm He}$ (5-504 $\rm \AA$) flux and stellar activity, and found that a higher XUV$_{\rm He}$ flux and a higher stellar activity contribute to the detectability of He 10830 excess absorption. Using similar comparing methods, \cite{2022A&A...659A..55O} and \cite{2023AJ....165...62Z} also showed that planets with a relatively larger XUV$_{\rm He}$ flux tend to give a He 10830 excess absorption.  \cite{2023SciA....9F8736Z} examined the plots of absorption vs. Roche-lobe filling, planets' surface gravities, equilibrium temperatures, incident XUV flux, and their host stars' effective temperatures in the exoplanets with detections and upper-limit constraints of He 10830.
However, they found no clear trend of He 10830 absorption with these parameters. \cite{2023A&A...677A.164A} searched for helium atmosphere in a sample of eleven exoplanets and showed the He 10830 absorption as a function of stellar mass and XUV$_{\rm He}$ flux. They suggested that the dependence of He 10830 absorption on XUV$_{\rm He}$ flux is influenced by other parameters, such as the stellar mass.

\section{Conclusions}\label{sec: consclusion}

In this work, we modelled the H$\alpha$ and He 10830 transmission spectra of HAT-P-32b simultaneously by using a self-consistent hydrodynamic model, which is coupled with a non-local thermodynamic model to calculate the level populations of H(2) and H(2$^3$S). A Monte Carlo simulation of Ly$\alpha$ resonance scattering was performed to calculate the Ly$\alpha$ radiation intensity inside the atmosphere, which is essential to calculate the population of H(2). We fitted the H$\alpha$ and He 10830 absorption lines with plausible assumptions about the stellar flux. We also provide a means to estimate the stellar Ly$\alpha$ flux, which is usually not directly observed. Although it is possible the heavy species can be dragged into the upper atmosphere of HAT-P-32b, using the constraints of stellar Ly$\alpha$ flux, we probed that the upper atmosphere of HAT-P-32b has not a super-solar metalicity, but a near solar metallicity, opposite to the lower atmosphere.

For fitting the He 10830 absorption line, relatively low $F_{\rm XUV}$ and $\beta_m$ are needed, while to explain the H$\alpha$ absorption, large $F_{\rm XUV}$ and/or $\beta_m$ are required. The H/He ratio does not significantly affect the H$\alpha$ absorption, especially when H/He $>$ 99/1. However, the He 10830 absorption strongly depends on H/He. In particular, for H/He = 92/8, only models with $F_{\rm XUV} = 0.1 F_0$ and $\beta_m = 0.1$ can explain the He 10830 observation. Large $F_{\rm XUV} > 3 F_0$ and $\beta_m \geq 0.3$ are required for the case of H/He = 99.9/0.1. Fitting both lines simultaneously, we constrained the hydrogen-to-helium abundance ratio to be H/He $\geq$ 99.5/0.5, the XUV flux to be approximately (0.5-3.0) times the fiducial value ($F_0 = 4.2 \times 10^{5}$ erg cm$^{-2}$ s$^{-1}$), and the spectral index $\beta_m$ to be about 0.16-0.3. The final models give a mass-loss rate of about (1.0-3.1)$\times 10^{13} \rm g \ s ^{-1}$ and a temperature of about 11,400-13,200 K.
Our results are consistent with the previous studies \citep{2022AA...657A...6C,2023A&A...673A.140L}. Moreover, our results show that the stellar Ly$\alpha$ flux can be as high as $4 \times 10^{5}$ erg cm$^{-2}$ s$^{-1}$, consistent with the high stellar activity at the observation epoch of the H$\alpha$ and He 10830 absorption.

In this paper, we investigated the transmission signals of H$\alpha$ and He 10830 observed by \cite{2022AA...657A...6C} at mid-transit. In the near future, we plan to study the time series of these transmission signals in order to obtain more accurate constraints on the upper atmosphere of this planet.
To study the escape of helium in more detail and find its statistical trends, more observations are needed. Advanced models both in 1D, which are faster and hence adequate for parameter studies, and sophisticated 2-3D models, which include more physical processes, are needed for a detailed explanation of the observed signals, especially the asymmetric transmission features.

\clearpage

\begin{table}[!hp]
    \centering
    \footnotesize
    \setlength{\tabcolsep}{6pt}
    \renewcommand{\arraystretch}{1.3}
\caption{Paramaters and model settings}    
   \begin{tabular}{lll}
     \hline
     \hline
 Parameters (the fiducial model) & Value & Reference \\
 \hline
$M_\star$ (stellar mass) & 1.160 $M_\sun$ &  \cite{2011ApJ...742...59H} \\
$R_\star$ (stellar radius) & 1.219 $R_\sun$ &  \cite{2011ApJ...742...59H}  \\
$M_P$ (planetary mass) & 0.585 $M_J$  &  \cite{2011ApJ...742...59H}  \\
$R_P$ (planetary radius) & 1.789 $R_J$ & \cite{2011ApJ...742...59H}   \\
Semi-major axis & 0.0343 AU & \cite{2011ApJ...742...59H}  \\
$\rm [Fe/H]$ & -0.04 &  \cite{2011ApJ...742...59H} \\
H/He & 92/8  & This work \\
$L_{\rm X}$ or $L(5-100\rm\AA) $ & $2.3\times 10^{29}$ ergs$^{-1}$  & \cite{2022AA...657A...6C} \\
$L(100-504\rm\AA) $ & $3.1^{+4.1}_{-1.4}\times 10^{29}$ erg s$^{-1}$  & \cite{2022AA...657A...6C} \\
$L_{\rm EUV}$ or $L(100-920\rm\AA) $ & $1.2^{+2.1}_{-0.7}\times 10^{30}$ erg s$^{-1}$  &  \cite{2022AA...657A...6C}\\
$F_{\rm NUV}$ or $F(912-3646 \rm\AA)$ at 0.0343 AU & $\sim 6\times 10^{7} \rm erg$ cm$^{-2}$ s$^{-1}$  &  \cite{2022AA...657A...6C}\\
$F_{\rm XUV}$ or $F(1-912\rm\AA)$ at 0.0343 AU, $F_0$ & $4.2\times 10^{5} \rm erg$ cm$^{-2}$ s$^{-1}$  &  \cite{2022AA...657A...6C} \\
$F_{\rm X}$ or $F(1-100\rm\AA)$ at 0.0343 AU, $F_{\rm X_0}$ &  6.95 $\times 10^4$ erg cm$^{-2}$ s$^{-1}$ & \cite{2022AA...657A...6C}\\
$\beta_m = F(1-100\rm\AA )/F(1-912 \rm\AA)$ & 0.16 &  \cite{2022AA...657A...6C}\\
Ly$\alpha$ flux at 1 AU ($\zeta$ Dor) &  46.5 erg cm$^{-2}$ s$^{-1}$  &  \cite{2013ApJ...766...69L} \\
Ly$\alpha$ flux at 0.0343 AU, $F_{\rm Ly\alpha_0}$ &  39,524 erg cm$^{-2}$ s$^{-1}$  &  \cite{2013ApJ...766...69L}, this work\\
Pressure at the bottom of atmosphere & 1 $\mu$bar & This work \\
Atmosphere upper boundary & 10 $R_P$ &  This work \\
 \hline
  Parameter range adopted in this work & &  \\
  \hline
 Z & 1, 10, 30, 50, 100, 200, 300  (10 $\leq$ Z $\leq$ 50 for discussion)\\
 H/He & 92/8, 99/1, 99.5/0.5, 99.9/0.1 \\
 $F_{\rm XUV}$/$F_0$ &  0.125, 0.25, 0.5, 1, 1.25, 2, \\ 
 & (0.1 for H/He = 92/8, 2.5 and 3 for H/He = 99.9/0.1)\\
 $\beta_m$  & 0.1, 0.16, 0.2, 0.3, 0.4, 0.5 \\
 $F_{\rm Ly\alpha}/F_{\rm Ly\alpha_0}$ & 1, 3, 5, 10, 30, 50, 100, 200, 300, 400, 500, 1000, 4000, 10,000, 20,000 \\ 
 \hline
   \end{tabular}
\label{tab:paramaters}
\end{table}

\clearpage

\begin{acknowledgements}
We thank the anonymous referees for their constructive
comments, which helped improve the manuscript.
This work is supported by the Strategic Priority Research Program of the Chinese Academy of Sciences, grant No. XDB 41000000, the National Key R\&D Program of China (grant No. 2021YFA1600400/2021YFA1600402), the National
Natural Science Foundation of China (grants Nos. 11973082, 12288102 and 42305136), the Natural Science Foundation of Yunnan Province (Nos. 202201AT070158, 202401CF070041), and the International Centre of Supernovae, Yunnan Key Laboratory (No. 202302AN360001).
K.-I. Seon was partly supported by a National Research Foundation of Korea (NRF) grant funded by the Korean government (MSIT; No. 2020R1A2C1005788) and by the Korea Astronomy and Space Science Institute grant funded by the Korea government (MSIT; No. 2023183000). SC acknowledges the support of the DFG priority program SPP 1992 “Exploring the Diversity of Extrasolar Planets" (CZ 222/5-1). The authors gratefully acknowledge the “PHOENIX Supercomputing Platform” jointly operated by the Binary Population Synthesis Group and the Stellar Astrophysics Group at Yunnan Observatories, Chinese Academy of Sciences. The IAA team acknowledges financial support from the Agencia Estatal de Investigaci\'on, MCIN/AEI/10.13039/501100011033, through grants PID2022-141216NB-I00 and CEX2021-001131-S.
  
\end{acknowledgements}

\bibliographystyle{a_a}
\bibliography{aa}

\end{document}